\documentclass[epj]{svjour}

\usepackage{graphicx}
\usepackage{epsfig}
\usepackage{dcolumn}
\usepackage{bm}
\def\lessim{\lower.5ex\hbox{$\; \buildrel < \over \sim \;$}}
\newcommand{\pathnow}{}
\begin{document}
\hbadness=10000
\topmargin -1.3cm\oddsidemargin = -0.1cm\evensidemargin = -1.3cm

\title{Hadron Production and Phase Changes\\ in Relativistic 
Heavy Ion Collisions}
\author{Jean Letessier\inst{1}\inst{2}  \and Johann Rafelski\inst{1}\inst{2}\inst{3}
}
\institute{Department of Physics, University of Arizona, Tucson, Arizona, 85721 USA, \and Laboratoire de Physique Th\'eorique et Hautes Energies
Universit\'e Paris 7, 2 place Jussieu, F--75251 Cedex 05, \and
CERN-PH-TH, 1211 Geneva 23, Switzerland}
\date{April 22, 2005; revised February 7  2008}

\abstract{
We study  soft  hadron production  
in relativistic heavy ion collisions in a wide range 
of reaction energy, $4.8\,{\rm GeV}<\sqrt{s_{\rm NN}}<200\,{\rm GeV}$,
and make predictions about yields of particles using the statistical
hadronization model. In fits to experimental data, we obtain
  both the statistical parameters
as well as physical properties of the hadron source. 
We identify the properties of the fireball at
the   critical energy threshold, 
$6.26\,{\rm GeV}<\sqrt{s_{\rm NN}^{\rm cr}}<7.61\,{\rm GeV}$,
 delineating  for higher energies hadronization of an 
entropy rich phase. In terms of the  chemical composition, 
one sees a phase which at low energy is chemically under-saturated, and 
which turns   into  a chemically over-saturated state persisting 
 up to the maximum accessible   energy. Assuming  that there
is no change in physical mechanisms in the energy range 
$15>\sqrt{s_{\rm NN}}\ge 200\,{\rm GeV}$, we use continuity of
particle yields and statistical parameters to predict
the hadron production  at $\sqrt{s_{\rm NN}}=62.4\,{\rm GeV}$,
and  obtain total yields of hadrons at  
$\sqrt{s_{\rm NN}}=130$\,GeV. We consider,
in depth, the pattern we uncover within the  hadronization condition, 
and discuss possible mechanisms associated with the identified 
rapid change in system properties at $\sqrt{s_{\rm NN}^{\rm cr}}$.
We propose that the chemically over-saturated 2+1 flavor  hadron matter system 
undergoes a 1st order phase transition. 
}

\maketitle
\section{Introduction} \label{intro}
\vskip -12.5cm
\hspace*{13cm} \underline{CERN-PH-TH/2005-060}\\
\vskip 11.6cm
It is   believed that the deconfined phase of matter  is
formed at sufficiently high energy and reaction  volume   
reached  in  the most central collisions of heavy ions at the 
 top RHIC energy~\cite{RBRC}. The question is where this critical 
energy threshold $\sqrt{s_{\rm NN}^{\rm cr}}$ is. 
We pursue this point in this systematic study, in order to explore  possible
phase changes occurring as function of collision energy~\cite{Gaz}. We 
furthermore 
compare our results qualitatively to the behavior seen as function of the 
reaction volume~\cite{RafRHIC}.

The tool, used in our study of soft hadron production, is the 
generalized statistical hadronization  model (SHM) which allows for 
particle yields to be in full chemical non-equilibrium~\cite{Letessier:1998sz}.
SHM is  capable to describe,  in detail, hadron  abundances and has 
been considerably refined in past decade, after its formulation 
by Fermi and Hagedorn~\cite{KMR03}.
 
We present and/or  extend here results of  analysis of 
 the energy dependence of total hadron  production  yields for:\\ 
a) fixed target symmetric   Au--Au reactions at the top 
available AGS   projectile energy  11.6 $A$GeV
(energy per colliding nucleon pair $\sqrt{s_{\rm NN}}=4.84$~GeV),\\
b) fixed target symmetric Pb--Pb reactions at 
SPS at 20, 30, 40, 80 and 158  $A$GeV  projectile energy. This 
we refer to as SPS energy range,  
$6.26\le \sqrt{s_{\rm NN}}\le 17.27$ GeV,\\ 
c) the Au--Au reactions in the collider mode 
at RHIC  in 65+65, 100+100, and also at 31.2+31.2  $A$GeV
reactions for  both total, and central rapidity yields.
This is the RHIC energy range,   $62.4\le \sqrt{s_{\rm NN}}\le 200$ GeV.

Experimental data analysis at RHIC was  carried out 
 for the central rapidity region yields 
at $\sqrt{s_{\rm NN}}=130$ and 200 GeV, and for the full hadron yields
at 200 GeV. The results we present for total hadron 
yields at $\sqrt{s_{\rm NN}}=130$
 GeV and 62.4 GeV,  arise from  our model considerations alone.
This is also in part the case for   for the central rapidity  yields at 
$\sqrt{s_{\rm NN}}= 62.4$ GeV .

As a first step, we aim to describe  at each reaction energy 
 the hadron yield data.  We obtain in this process the 
 statistical hadronization model (SHM) parameters, which 
allow to evaluate the yields of  all (also of unobserved) 
particles.  One can see
SHM analysis  as a method of how   the known experimental hadron yield 
data can be extrapolated to obtain   the unobserved  hadron yields.
For this reason, we  also attempt to extrapolate to reaction
energies and phase space coverage which is not allowing, for 
lack of data, a SHM fit. For example, for the 31.2+31.2 $A$GeV   case, 
we interpolate  strange particle yields, which are known below and above this 
energy, and/or fix   certain SHM parameters which show continuity 
as function of reaction energy, respecting in the process the constraints 
of the SHM. 

In our analysis, we are  seeking consistency in the results across
the reaction energy. This is  of importance when the number of measurements
is not much greater than is the number of observables. When we are able to fix
the values of statistical parameters by consistency check  across energy
range, then the number of used parameters is reduced and the 
statistical significance shown in this work is for the number of parameters 
actually fitted. However, more often,  in the study of
statistical significance,   we do not account explicitly 
for consistency across energy range. For this reason, in most 
cases, the statistical significance we present  is a lower limit.  

Once a statistically significant description of the data sample at an
energy is achieved, we have available  the yields of all soft
 hadronic particles and their resonances.  We  sum 
 partial contributions of each particle species to
 quantities such as  entropy, strangeness, baryon number,
to obtain the  properties of the fireball at the time of particle
 production (hadronization).   In this way,  we evaluate fireball  breakup
pressure,  entropy, baryon number, strangeness yield  and 
the thermal energy content. In this approach,   
the  kinetic energy content  associated with the collective  flow  
of matter is not considered --- this requires a study of particle
$m_\bot$ and rapidity spectra, beyond the integrated  hadron yields. 

High strangeness~\cite{Rafelski:1982pu}, and entropy
 content~\cite{Glendenning:1984ta,Letessier:1992xd} 
of a dense hadronic matter fireball are  the anticipated 
characteristic property of the color deconfined state of matter.
Once formed, this  enhancement of strangeness and entropy
is also the  property of  the final hadronic state: first principles 
require that  entropy must increase in the fireball 
expansion, as well as in  the ensuing  hadronization process; model studies
show that once strangeness is produced, it  remains present during
expansion of dense matter, it can slightly increase during 
 hadronization~\cite{KMR,Koch:1984tz}.  

Particle yields, and pion yield  in 
particular, provide natural measure of entropy yield, while the kaon yields, 
and in particular the ${\rm K^+}$ yield, are an approximate
 measure of the total strangeness yield across all 
reaction energies~\cite{Glendenning:1984ta}. 
The yield ratio  ${\rm K^+}/\pi^+$ has been 
studied as function of reaction energy in the SPS energy domain
 and a strong `horn' like
feature has been discovered~\cite{Gaz}.  

This suggests a   change in the reaction 
mechanism of particle production,  occurring in central collisions
of Pb--Pb, in the energy interval
$6.26\,{\rm GeV}\le\sqrt{s_{\rm NN}}\le 7.61\,{\rm GeV}$, the 
two limits correspond to 20 and respectively, 30 $A$GeV Pb beams 
on fixed target.   This energy range is just at the predicted threshold of quark--gluon 
formation arising  considering balance of energy deposition and relativistic 
reaction dynamics~\cite{DanRaf}. Possibility of a rich phase structure of
the deconfined phase at high baryochemical potential and finite temperature
further enhances our interest in the study of this reaction energy domain~\cite{Kitazawa:2001ft}. 

To describe  experimental results indicating the presence of a critical (`cr') energy 
threshold, one can, in first instance, use two different reaction models 
which apply below and, respectively, above  a postulated energy 
threshold for a phase transformation~\cite{GazHorn}. However, 
this presupposes the most important outcome, namely that there is an 
energy dependent change in dense matter fireball structure at its breakup. 
Moreover,  such an approach does not produce as result of analysis an  
insight into the structural change that occurs, and which could be
compared with predictions. Instead, the structural change is 
part of the hypothesis  under which the analysis is carried out.
For this reason, the  methodology  we choose here is more general.

We use in this work the software package  SHARE (statistical hadronization with
resonances)~\cite{share}, the public SHM suit of programs,  
where the methods of SHM analysis are described in greater detail.
Of particular importance here is that the full mass  spectrum of hadron resonances 
is included~\cite{Torrieri:2003nh}.
SHARE implements two  features  important for the full understanding
of the  ${\rm K^+}/\pi^+$ horn:\\
1) the isospin asymmetry driven 
by proton--neutron  asymmetry,   which is particularly relevant
 at low reaction energies,\\ 
2) the chemical  non-equilibrium (phase space under-satu\-ration
 and over-saturation) for strange and light quarks.\\ 
These two features appear to be  essential to
obtain a description of the   ${\rm K}^+/\pi^+$ energy dependent 
yield.  

We first describe, in next section \ref{Data},
  features of the data sample we use, discuss 
the input data and results of the fits for 
the AGS/SPS and RHIC energy range separately. 
We discuss the  
resulting  statistical  parameters and  
the confidence level of our fits. We survey, in both 
tabular and  graphic form, the   energy dependence of
particle yields  of interest, including an explanation
of the  ${\rm K^+}/\pi^+$ horn.   We then discuss   physical  
properties of the fireball at point of chemical freeze-out
in section \ref{Fire}, show the energy dependence of the model
parameters and of physical properties, and address the 
 strangeness and entropy production.
We discuss the results of our analysis and present their interpretation
in the final section \ref{results}. 

\section{Fit procedure and   hadron  multiplicities}\label{Data}
\subsection{General remarks}\label{general}
The measured experimental results are 
available  for either total particle 
yields, $N_{4\pi}$, or for central rapidity 
yields,   $dN/dy$. At RHIC energy scale,
we will study both data sets, though  $N_{4\pi}$
is rather incomplete at this time. At AGS and SPS, we will
solely consider  $N_{4\pi}$, in order to minimize the impact
of the shape of the  longitudinal unstopped matter flow
on the outcome of the analysis. 

At SPS, a  semi-distinct  central rapidity 
domain is only present in the top SPS case,   its re-analysis
will make good sense once the RHIC   $dN/dy$ data extend to 
the minimum accessible energy  domain which is close to 
top SPS energy range. However, this will  require introduction
of models of collective matter flow, a step which we 
do not wish to take in this work. At high RHIC energies, we
presume that the fragmentation regions are sufficiently 
separated from the central rapidity domain as to allow 
the study of the rapidity particle distributions $dN/dy$, 
at central rapidity, in a model independent fashion. 

 We include, in our consideration 
of the total particle yields $N_{4\pi}$, the 
trigger condition which defines the 
participant `wounded' nucleon number $N_W$.
This has to be equal to the total net baryon number $b=B-\overline B$ 
contained within the final state  particle multiplicities.
Furthermore, both for   $N_{4\pi}$  and  
 central rapidity yields $dN/dy$,  we
consider two constraints: \\
a) the fraction of 
protons among all nucleons (0.39 for heavy nuclei)  establishes a fixed 
final  ratio of all  electrical
charge $Q$ to the total final state baryon number $b$ -- the 
ratio $Q/b$ is preserved in any fraction of a volume of centrally colliding nuclei; 
it is a measured quantity given  the fact that both target and projectile are known;\\
b) strangeness (valance $s$-quarks) 
content of hadrons prior to weak decays has to be
(up to systematic experimental error)
 balanced by antistrangeness (valance $\bar s$-quarks) 
 bound in hadrons  for the  $N_{4\pi}$
study, and nearly balanced  
when considering the central rapidity $dN/dy$ distributions. 
 
As our prior studies showed~\cite{IJMPE}, any deviation from strangeness
 conservation as function of rapidity is, in general, 
smaller than the typical 10\% systematic error 
of the experimental data points . It is the level of systematic error  
in the particle yields
which determines the  precision at which  we have to assure  
strangeness conservation. Forcing exact balance 
can create an aberration of the fit, since the sharp  constraint 
is inconsistent with several independent measurements which
contribute to the cancellation. For example,  
 at several SPS energies the systematic errors between 
${\rm K}^+$ and $\Lambda$ which control  the yields of $\bar s$ and,
respectively, $ s$ quarks, do not cancel to better than 8\% level. 
 This can be  checked without a fit in a qualitative study of 
the key particle  yields.   

Another reason to be cautious with the strangeness conservation is that 
  the spectra of hadrons we are using could contain 
wrong entries (e.g., pentaquark states which we in view of 
recent experimental results do not anymore include in the input 
data set, or  wrong spin-isospin assignments for little known 
states). Moreover, we maybe    missing some relevant undiscovered 
resonances. These effects 
are largest when the baryon asymmetry is largest, since the 
strangeness balance condition probes  at large baryochemical
potential the mass spectrum of strange baryons and mesons 
separately, with mesons dominating in antistrangeness and baryons
important in the strangeness count. 

For this reason, our strangeness conservation   procedure is 
as follows: when a first fit shows strangeness  a slight 
strangeness asymmetry, we   find the best  parameters for
the fit with a loose, systematic error related
 strangeness conservation constraint allowing, 
e.g., a  10\% deviation from balance as a fit input
that is we request $(s-\bar s)/(s+\bar s)=0\pm0.1$. 
Since we present confidence level and profiles of the fit, 
and we wish to have from energy to energy comparable results,
 we  redo the fit with a  fixed   preferred value of 
the  strangeness fugacity $\lambda_s$  as
is done in case of using exact strangeness conservation.

In this way, we obtain a data fit with
the same mechanism of approach as for the  cases where
exact strangeness conservation is  used to fix 
one parameter, so that confidence levels are comparable. 
 We find $ -0.07> (s-\bar s)/(s+\bar s)> -0.1$ in the  SPS energy 
domain.  The asymmetry   favors 
an over-count of $\bar s$ quarks in emitted hadrons. It  is moderate
in its relative magnitude, staying within the systematic errors of
the measurements used  in this study. We will state the strangeness
balance explicitly when presenting the computed particle yields.
Note that  addition of   pentaquarks to the hadron spectrum decreases 
this asymmetry by 0--3\%, but has otherwise minimal influence on 
the fit results presented. 

As the above discussion of strangeness conservation shows,
conserved quark quantum numbers introduce  yield
constraints, which particle 
multiplicities  cannot deviate from. How a subset
of SHM  parameters
determine a set of  particle ratios has been shown for
the first time in 1982~\cite{Koch:1982ij}.  An nice example
is the chemical relation between the ${\rm K}^-/{\rm K}^+$ and $\bar p/p$ 
 demonstrated experimentally in 2003, see figure 4 in~\cite{Bearden:2003fw},
a development  based on the rediscovery of the   SHM
constraints in 2000~\cite{Becattini:2000jw}.  
Since  SHM with its  chemical consistencies 
has been  very successful in helping understand hadron
production, we embark on  further data verifications  
at each energy,  checking the  
consistency of experimental  data with SHM.  

 A suspect particle yield   can be further cross checked 
studying  the    behavior of this  particle yield  
as  function of energy. Such consideration is very 
important since we are searching  for a change in the
physical properties of the fireball as function of energy, 
and we do not want the outcome to be even in part the
result of a statistical fluctuation
in the reported yield of a subset of  particles. 
 We find inconsistencies (see next paragraph) in the particle yield
effects. None of these influence decisively 
 the findings we report here, in part because of the more
lax attitude we take toward the constraint on strangeness
conservation we described above. Moreover, 
 considering the large number of experimental data considered, fluctuations 
 in experimental data sample 
must occur.
 
Specifically,  we did not use the 
 $\Lambda(1520)$ nor $\Omega$ and $\overline\Omega$ yields
obtained at 158 $A$ GeV in our fit. The   preliminary  $\Lambda(1520)$
value at top SPS  is   $\Lambda(1520)=1.45\pm0.4$~\cite{Friese:2002re}. 
This is within 3 s.d. of the SHM  yield. However, this exceptionally 
narrow resonance may be subject to additional effects~\cite{Rafelski:2001hp} 
and we felt that it is more prudent to not include its study here. 
The experimental yields of  $\Omega$ and $\overline\Omega$ at 158 $A$ GeV
are contrary to the $\Lambda(1520)$   larger than  the SHM model predicts.
These particles are produced very rarely and for this reason any novel 
mechanism of production~\cite{Kapusta:2000ny}  would be first visible in 
their yield. We believe that it is also prudent to not include these
in the study, even if the deviation from fit would be at 2 s.d. level.

\subsection{AGS and SPS energy range fit}\label{ASfit}
To assure the reproducibility of our analysis, we will describe 
in detail the input particle yields that are used, 
for AGS/SPS energy domain, and for RHIC domain in the next subsection. 
The set of particles available at AGS arises from several experiments,
we have previously reported in detail the SHM analysis at the top
 AGS energy~\cite{Letessier:2004cs}, which input and fit 
results are restated here. 

The study of AGS results was performed in~\cite{Letessier:2004cs} 
for several possible cases, such as with and without $\phi/K$ yield, 
strangeness (non)conservation. The results here presented are for the
case in which the $\phi/K$ yield is fitted and strangeness is conserved.
Differences in theoretical fit detail yield similar fit result which
show the robustness of the approach.

For this work,  the analysis of   the  $N_{4\pi}$ 
particle yields of the  NA49 experimental
 group available at 20, 30, 40, 80 and 158 $A$ GeV \cite{GazPriv} 
has been carried out. This work   extends significantly our prior study 
of the  40, 80 and 158 $A$ GeV NA49 done when many 
fewer  measurements  were available~\cite{Fit03}.
Moreover, the SHARE package used offers additional 
theoretical features which were 
not fully implemented  earlier: the consistent description of the yields of
 different charges hadrons (e.g., protons and neutrons, $\pi^+$ and $\pi^-$
by means of  $\lambda_{I3}$ allows to fix the
net charge fraction $Q/b$.  The  most relevant difference
to the earlier study is, however, 
 that we can address the two newly measured reaction energies, 
 20, 30 $A$ GeV. This, along with the AGS 11.6 $A$ GeV data, including  the 
recently published   $\phi$-yield~\cite{AGSphi}, 
allows to recognize  a major change in the behavior 
of the hadronizing fireball~\cite{Gaz}.

\begin{table*}[!t]
\caption{
\label{AGSSPS}
The input $N_{4\pi}$ total particle multiplicities data at top, and, below, 
the resulting statistical parameters  for AGS and SPS energy range.   At bottom,
we state the chemical potential corresponding to the central values of the fugacities.  
For each projectile energy $E$ [$A$GeV], we also present in the header the
invariant center of momentum
 energy per nucleon pair, $\sqrt{s_{\rm NN}}$  [GeV], the center of
momentum rapidity and the centrality of the reaction considered. 
The $\lambda_s$  values  marked with
a $^*$ are  result of a strangeness conservation constraint.
}\vspace*{0.2cm}
\begin{center}
\begin{tabular}{|c| c | c c c c c |  }
\hline
E[$A$GeV]                      & 11.6          & 20          & 30              & 40         & 80            & 158  \\
$\sqrt{s_{\rm NN}}$  [GeV]     &4.84           &6.26         &7.61             &8.76          &12.32          &17.27  \\
$y_{\rm CM}$                   &1.6  &1.88 &2.08 &2.22 & 2.57& 2.91 \\
\hline
$N_{4\pi}$ centrality          &most central   &  7\%          &  7\%          &  7\%           &  7\%           &  5\%\\
\hline
$R=p/\pi^+$,  $N_W $
                        &$R=1.23 \pm 0.13$  &349$\pm$6      &349$\pm$6      &349$\pm$6      &349$\pm$6      &362$\pm$6           \\
$Q/b $                      &0.39$\pm$0.02  &0.394$\pm$0.02 &0.394$\pm$0.02 &0.394$\pm$0.02 &0.394$\pm$0.02 &0.39$\pm$0.02   \\
$\pi^+$                     &133.7$\pm$9.9  &184.5$\pm$13.6 &239$\pm$17.7   &293$\pm$18     &446$\pm$27     &619$\pm$48       \\
$R=\pi^-$ /$\pi^+$, $\pi^-$
                        & $R=1.23 \pm 0.07$ &217.5$\pm$15.6   &275$\pm$19.7 &322$\pm$19     & 474$\pm$28     &639$\pm$48       \\
$R={\rm K}^+/{\rm K}^-$, ${\rm K}^+$   
                            &$R=5.23\pm0.5$  &40$\pm$2.8     &55.3$\pm$4.4   &59.1$\pm$4.9   &76.9$\pm$6     &103$\pm$10      \\
${\rm K}^-$                 &3.76$\pm$0.47  &10.4$\pm$0.62  &16.1$\pm$1     &19.2$\pm$1.5   &32.4$\pm$2.2   &51.9$\pm$4.9      \\
$R=\phi/{\rm K}^+$, $\phi $ 
                         &$R=0.025\pm 0.006$&1.91$\pm$0.45  &1.65$\pm$0.5   &2.5$\pm$0.25   &4.58$\pm$0.2   & 7.6$\pm$1.1      \\
$\Lambda$                   &18.1$\pm$1.9   &28$\pm$1.5     &41.9$\pm$6.1   &43.0$\pm$5.3   &44.7$\pm$6.0   &44.9$\pm$8.9       \\
$\overline\Lambda$          &0.017$\pm$0.005&0.16$\pm$0.03  &0.50$\pm$0.04  &0.66$\pm$0.1   &2.02$\pm$0.45  &3.68$\pm$0.55    \\
$\Xi^-$                     &               & 1.5$\pm$0.13  & 2.48$\pm$0.19 &2.41$\pm$0.39  &3.8$\pm$0.260  & 4.5$\pm$0.20      \\
$\overline\Xi^+$            &               &               &0.12$\pm$0.06  & 0.13$\pm$0.04 &0.58 $\pm$0.13 & 0.83$\pm$0.04      \\
$\Omega+\overline\Omega$    &               &               &               &0.14$\pm$0.07   &              &                   \\
${\rm K}_{\rm S}$           &               &               &               &                &              & 81$\pm$4           \\
 \hline 
$V {\rm [fm}^3]$            &3596$\pm$331   &4519$\pm$261   &1894$\pm$409   &1879$\pm$183    &2102$\pm$53   & 3004$\pm $1  \\
$T$ [MeV]                   &157.8$\pm$0.7  &153.4$\pm$1.6  &123.5$\pm$3    &129.5$\pm$3.4   &136.4$\pm$0.1 &136.4$\pm$0.1       \\
$\lambda_q$                 &5.23$\pm$0.07  &3.49$\pm$0.08  &2.82$\pm$0.08  &2.42$\pm$0.10   &1.94$\pm$0.01 & 1.74$\pm$0.02     \\
$\lambda_s$                 &1.657$^*$      &1.41$^*$       &1.36$^*$       &1.30$^*$        &1.22$^*$      & 1.16$^*$   \\
$\gamma_q$                 &0.335$\pm$0.006  &0.48$\pm$0.05  &1.66$\pm$0.10  &1.64$\pm$0.04  &1.64$\pm$0.01 &1.64$\pm$0.001\\
$\gamma_s$                  &0.190$\pm$0.009 &0.38$\pm$0.05  &1.84$\pm$0.32  &1.54$\pm$0.15  &1.54$\pm$0.05 & 1.61$\pm$0.02     \\
$\lambda_{I3}$             &0.877$\pm$0.116  &0.863$\pm$0.08 &0.939$\pm$0.023&0.951$\pm$0.008&0.973$\pm$0.002& 0.975$\pm$0.004     \\
\hline
$\mu_{\rm B}$ [MeV]               &783      &576            &384            &344             &271           & 227   \\
$\mu_{\rm S}$ [MeV]               &188      &139            & 90.4          &80.8            &63.1          & 55.9  \\
 \hline
\end{tabular}\vspace*{0.1cm}
\end{center}
 \end{table*}

 The input data we considered for the  AGS and SPS 
 are   presented   in top part of the  table \ref{AGSSPS}. The statistical
parameters are seen below these input data.  
In carrying out the data analysis,  we  use the full grand-canonical
statistical set of seven  parameters: volume V, freeze-out temperature $T$,
chemical quark fugacities $\lambda_{q,s}$, 
quark occupancy parameters $\gamma_q$ and $\gamma_s$, and third component of the 
isospin fugacity $\lambda_{\rm I3}$. The fitted values of these 7 parameters 
are seen near bottom of the table \ref{AGSSPS}, which is
followed by entries for the central values of the two chemical  potentials:
\begin{eqnarray}
\mu_{\rm B}&=&3T\ln \lambda_q,\\
\mu_{\rm S}&=&\mu_{\rm B}/3-T\ln \lambda_s.
\end{eqnarray}   

The uncertainties in the value of statistical parameters comprise the 
propagation of  experimental measurement error through the fit, as well as 
ambiguity due to statistical parameter correlations arising. In some instances
this effect is very small, in others rather large. This wide disparity  is possible, 
as sometimes the data set is sufficiently constraining, and in others it is not. 
The most interesting result, we notice in   table \ref{AGSSPS},
 is the sudden shift in the values
of the phase space occupancies $\gamma_q$ and $\gamma_s$  observed as 
reaction energy rises from 20 to 30 $A$\,GeV.
The value of chemical freeze-out temperature $T$
 changes accordingly to counterbalance the
effect of a rapid change in  $\gamma_q$ and $\gamma_s$
 on some particle multiplicities. We will discuss
this change in behavior in great detail 
in what follows. The steady decrease  of 
baryochemical  potential  $\mu_{\rm B}$ with 
reaction energy follows  the enhancement in global yield
of hadrons.  At central rapidity  the  steady increase of   
baryon transparency  with increasing collision energy 
yields a smaller value of  $\mu_{\rm B}$.  The  
total particle yields we consider here yield an average over
the entire rapidity range of  $\mu_{\rm B}$. 
The associated value of $\mu_{\rm S}$ is controlled by 
strangeness conservation condition, as discussed.
 
\begin{figure}[!t]
\hspace*{-.20cm}\epsfig{width=9cm,figure=\pathnow   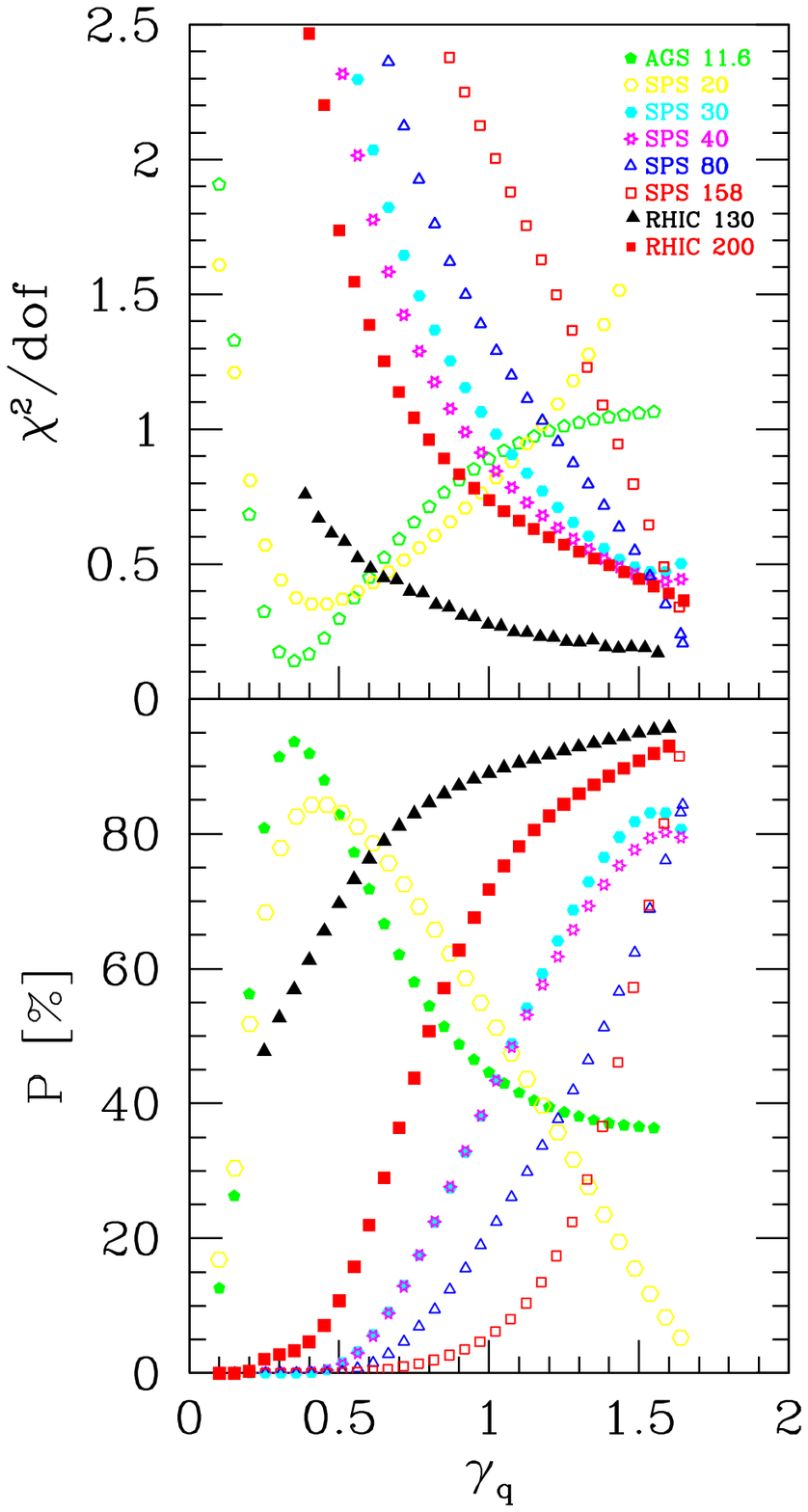}
\caption{\label{ChiP}
$\chi^2/{\rm dof}$ (top) and the associated  significance level  $P[\%]$
(bottom) as function of $\gamma_q$, the light quark phase space occupancy.
for the AGS/SPS energy range and for the (central rapidity) RHIC results.
}
\end{figure}

As seen in   table \ref{AGSSPS},
we occasionally fix the value of $\gamma_q$. The value we choose is the
the best value which emerges from study of $\chi^2$ profile, see figure  \ref{ChiP}.
We fix the best  $\gamma_q$ in order to reduce the correlations between 
parameters, given the  small number of degrees of freedom.
Excluding from the count of parameters $\lambda_s$ which is related
to (near)  strangeness conservation, there are 6
 (or 5 when $\gamma_q$ is fixed)  parameters
while there  are between 9 and 12 data inputs 
 at each energy considered
in  table \ref{AGSSPS}. We thus have  3--6 degrees
 of freedom (d.o.f) for the fits carried out at AGS and  SPS. 
Not all of the NA49 SPS energy range 
results we use  are published in final form. 

 We show, in figure \ref{ChiP},   the reliability of the  fits we obtained 
at different reaction energies as function of   $\gamma_q$, 
the light quark phase space occupancy. The results for AGS and SPS
are accompanied by those for central rapidity RHIC fits we will address below. 
The top frame, in figure~\ref{ChiP}, shows  
$\chi^2/{\rm dof}$. The associated  significance level  $P[\%]$
is seen in the bottom frame. 
We include $P[\%]$ as result, since the number of degrees of freedom in
each fit is small and it is hard to judge the significance of a 
small value of  $\chi^2/{\rm dof}$. 

We study dependence of $\chi^2/{\rm dof}$ and $P[\%]$ 
 on  $\gamma_q$ since we see, in table \ref{AGSSPS}, that  the two parameters 
which undergo  a rapid change as function of reaction  energy
are  $\gamma_q$, and to a lesser degree, the freeze-out temperature $T$.
The rapid change with  $\gamma_q$  is prominent  in  figure \ref{ChiP} where 
  $P[\%]$ peaks for the lowest
two energies (11.6 and 20 $A$GeV) at $\gamma_q<0.5$, while for
all other collision energies it grows to maximum value near
  $\gamma_q\simeq 1.6$, where  the Bose  singularity of the 
pion momentum distribution $\gamma_q\simeq e^{m_\pi/T}$. 

The  reader  can see, in  figure \ref{ChiP},   that setting the value 
$\gamma_q=1$ will yield
a set of energy dependent {\it individual} fits which appear
to have a good confidence level. However, the energy dependence of 
the particle yields derived at this fixed $\gamma_q=1$  condition 
is less  convincing. It is the rapid shift in 
the best   $\gamma_q$  as function of reaction energy which allows to
describe the `horn' feature in  the ${\rm K}^+/\pi^+$
data (see below figure~\ref{KPi}). Without  variable 
 $\gamma_i$  this horn feature is
largely erased,  see, e.g., figure 4 in \cite{Cleymans:2004hj}, 
and the dashed and dotted lines in figure~\ref{KPi}. We will 
return to describe this effect in section \ref{energy}, and 
discuss this physics in more detail in section \ref{horn}.
We believe that, in the study of energy dependent particle yields, 
 the use of the highest confidence level SHM
 results with  $\gamma_q\ne 1$  is required for
 the description of energy dependent particle yield data.

\begin{table*}[!t]
\caption{
\label{outputAGSPS}
Output total hadron multiplicity data   for AGS (left) and SPS (right).
Additional significant digits are presented in particle yields
for purposes of tests and verification.
The statistical parameters generating these multiplicities   are 
the central fit values seen in table \ref{AGSSPS}.
Hadron yields presented are prior to weak decays and apply to 
 the total multiplicities $N_{4\pi}$ expected at the most central collision 
bin with the corresponding baryon content $b$ as shown.
For SHM parameters, see table \ref{AGSSPS}.  
}
\vspace*{0.2cm}
\begin{center}
\begin{tabular}{|c| c | c c c c c | }
\hline
$E$ [$A$\,GeV]             &11.6 & 20  & 30    & 40    & 80  & 158  \\
$\sqrt{s_{\rm NN}}$  [GeV] &4.84 &6.26 &7.61   &8.76   &12.32 &17.27  \\
$y_{\rm CM}$               &1.6  &1.88 &2.08   &2.22   & 2.57& 2.91 \\
\hline
$N_{4\pi}$ centrality      &m.c. &  7\% &  7\% &  7\% &  7\% &  5\%\\
\hline
$b\equiv B-\overline B$    & 375.6&347.9 &349.2 &349.9 &350.3 &362.0 \\
$\pi^+$                    & 135.2&181.5 &238.7 &290.0 &424.5 &585.2 \\
$\pi^-$                    & 162.1&218.9 &278.1 &326.0 &461.3 &643.9 \\
${\rm K}^+$                & 17.2 &39.4  &55.2  &56.7  &77.1  &109.7 \\
${\rm K}^-$                & 3.58 &10.4  &15.7  &19.6  &35.1  &54.1  \\
${\rm K}_{\rm S}$          & 10.7 &25.5  &35.5  &37.9  &55.1  &80.2  \\
$\phi $                    & 0.46 &1.86  &2.28  &2.57  &4.63  &7.25  \\
$p $                       &174.6 &161.6 &166.2 &138.8 &138.8 &144.3 \\
$\bar p$                   &0.021 &0.213 &0.68  &0.76  &2.78  &5.46  \\
$\Lambda$                  & 18.2 &29.7  &39.4  &34.9  &42.2  &48.3  \\
$\overline\Lambda$         &0.016 &0.16  &0.51  &0.63  &2.06  &4.03  \\
$\Xi^-$                    & 0.47 &1.37  & 2.44 &2.43  &3.56  &4.49  \\
$\overline\Xi^+$           &0.0026&0.027 &0.089 &0.143  &0.42  &0.82  \\
$\Omega$                   & 0.013&0.068 &0.14  &0.144 & 0.27 &0.38  \\
$\overline\Omega$          &0.0008&0.0086&0.022 &0.030 & 0.083&0.16  \\
${\rm K}^0(892)  $         & 5.42 &13.7 &11.03  &12.4  &18.7  &19.1 \\
$\Delta^{0} $              & 38.7 &33.43 & 25.02 &26.6  &27.2  &28.2  \\
$\Delta^{++} $             & 30.6 &25.62 & 22.22 &24.2  &25.9  &26.9  \\
$\Lambda(1520)$            & 1.36 &2.06  & 1.73 &1.96  &2.62  &2.99  \\
$\Sigma^-(1385)$           & 2.51 &3.99  & 4.08 &4.26  &5.24  &5.98  \\
$\Xi^0(1530) $             & 0.16 & 0.44 & 0.69 &0.73  &1.14  &1.44  \\
$\eta $                    & 8.70 &16.7  & 19.9 &24.1  &38.0  &55.2  \\
$\eta' $                   & 0.44 &1.14  & 1.10 &1.41  &2.52  &3.76  \\
$\rho^0 $                  & 12.0 &19.4  & 14.0 &18.4  &32.1  &42.3  \\
$\omega(782) $             & 6.10 &13.0  & 10.8 &15.7  &27.0  &38.5  \\
$f_0(980)$                 & 0.56 &1.18 & 0.83 &1.27  &2.27  &3.26  \\
\hline
$(s-\bar s)/(s+\bar s)$    & 0    &-0.092&-0.085 &-0.056&-0.029&-0.056  \\
\hline
\end{tabular}\vspace*{0.1cm}
\end{center}
 \end{table*}

Regarding the weak decay (WD) contributions: 
in the fits  to experimental data,
we have followed   the corrections
applied to the data by the experimental group(s).    
For  20 and 30 GeV in  $\Lambda$ and $\overline\Lambda$,
the data we use includes the WD of $\Xi,\ \overline\Xi,\ \Omega$ and $\overline\Omega$,
 these matter mainly in consideration of antihyperon yields.  
At all higher SPS energies all hyperon WD decays are corrected for
by the NA49 collaboration, the $\Sigma^\pm$ decays are always corrected. Similarly,  
decays of kaons into pions are corrected for at   all  SPS  energies.
 At AGS 11 GeV, all yields we consider are without WD contribution. 
The contamination of $\Lambda$ by hyperons decays  is not material. 
However, the decays of anti-hyperons contaminates in highly relevant 
way the yields of $\bar p$ and we do not discuss here this effect
further, the reader will note the relevant yields of $\bar p,\ \overline\Lambda$
and $\overline\Xi$ in table  \ref{outputAGSPS}. The observable  yield of $\bar p$ 
is further contaminated by decays of $\overline\Sigma^+$.

 The model  yields obtained are shown in table  \ref{outputAGSPS}.
These results are prior to any WD contributions.
The yields of input   particles  can 
be compared to the fitted inputs seen in table \ref{AGSSPS}. We 
present   also  predictions for yields of a number of other particles of   
interest. We do not show the uncertainty in these results, 
which can be  considerable: in addition to the error propagating 
through the fit, there is  systematic error due to the shape of the 
$\chi^2/{\rm d.o.f.}$ minimum,   see   figure \ref{ChiP}.

\subsection{RHIC energy range fit}\label{RHfit}
The RHIC central rapidity particle yields
 at $\sqrt{s_{\rm NN}}=200$ and 130 GeV  are  analyzed using 
nearly the  same method and principles   described  in the study of 
the  total particle yields. This can be done 
for the case that the particle yields, and hence 
their  source,  is subject to (approximate) scaling, that means 
is flat as function of the 
rapidity distribution~\cite{Cleymans:1999st}. The
overall normalization of yields then contains, 
instead of the volume $V$, the volume fraction $dV/dy$
associated with the size of the volume at the 
rapidity of the source of particles at $y$. We note that,
in the  local rest-frame, the total yield of particles $N_{4\pi}$
can be written in the  equivalent forms:
\begin{equation}\label{dNdV}
 N_{4\pi}=\int  dV \rho =
          \int  dy \frac{dV}{dy}\rho  = 
          \int  dy \frac{dN}{dy}.
\end{equation}

\begin{table*}[!t]
\caption{
\label{Rhic}
The input particle data (top) and 
the resulting statistical parameters,  and
the chemical potentials derived from these, at bottom,
for the RHIC energy range.  Any of the entries with 
a $^*$  is set as input or  is a constraint, e.g.,  in general $\lambda_s$ 
results from the  constraint
to zero strangeness. $^\dagger$ indicates input particle multiplicity derived from interpolating 
yields between different energies, see  the RHIC $\sqrt{s_{\rm NN}}=62.4$ GeV case.
On right, the case of central rapidity yields $dN/dy$, and on left, 
the total particle yields, in all
cases considered for the most central 7\%  collisions. For $N_{4\pi}$,  we show the  
 participant  count.
}\vspace*{0.2cm}
\begin{center}
\begin{tabular}{|c| c  c c | c c c |  }
\hline
$\sqrt{s_{\rm NN}}$  [GeV]  & 62.4         & 130            &200           & 62.4          & 130              & 200  \\
$E_{\rm eq}$ [$A$\,GeV]     &2075 & 9008  & 21321 &2075 & 9008  &  21321 \\
$\Delta y$ &$\pm 4.2$&$\pm 4.93$&$\pm 5.36$ &$\pm 4.2$  & $\pm 4.93$ & $\pm 5.36$ \\
\hline
 &  \multicolumn{3}{c}{  $N_{4\pi}$\ 5\% } & \multicolumn{3}{|c|}{ $dN/dy|_{y=0}$\ 5\% }\\
 \hline
$N_W $                      &349$\pm$6      &349$\pm$6      &349$\pm$6      &               &                 &              \\
$Q/b $                      &0.39$\pm$0.02  &0.39$\pm$0.02  &0.39$\pm$0.02  &0.39$\pm$0.02  &0.4$\pm$0.01     &0.4$\pm$0.01 \\
$\pi^-$ /$\pi^+$            &               &               &               &1.02$\pm$0.03 &1.0$\pm$0.03     &1.0$\pm$0.05 \\
$\pi^+$     &$^\dagger$1140$\pm$90    &$^\dagger$1450$\pm$90&1677$\pm$150   &               & 276$\pm$36      & 286.4$\pm$24.2   \\
$\pi^-$                     &               &               &1695$\pm$150   &               &270$\pm$36       &281.8$\pm$22.8  \\
${\rm K}^+$                 &               &               &293$\pm$26     &               &46.7$\pm$8       &48.9$\pm$6.3      \\
${\rm K}^-$                 &               &               &243$\pm$22     &               &40.5$\pm$7       &45.7$\pm$5.2  \\
$\phi/{\rm K}^-$            &               &               &               &               &0.15$\pm$0.03    &0.16$\pm$0.03 \\
$p$                         &               &               &               &               &28.7$\pm$4       &18.3$\pm$2.6  \\
$\bar p$                    &               &               &               &               &20.1$\pm$2.8     &13.5$\pm$1.8  \\
$\Lambda$                   &               &               &           & $^\dagger$17$\pm$2 &17.35$\pm$0.8    &                 \\
$\overline\Lambda$          &               &               &           & $^\dagger$10$\pm$1 &12.5$\pm$0.8     &                 \\
$\Xi^-$/$h^-$               &               &               &               &               &0.0077$\pm$0.0016 &                 \\
$\Xi^-$, $\Xi^-$/$\Lambda$  &               &               &           &$^\dagger$2.05$\pm$0.2&0.187$\pm$0.046&2.17$\pm$0.25                 \\
$\overline\Xi^+$, $\overline\Xi^+$/$\overline\Lambda$&      &       &   & $^\dagger$1.3$\pm$1  &0.215$\pm$0.054&1.83$\pm$0.25                 \\
$\overline\Xi^+$/$\Xi^-$    &               &               &               &               &0.853$\pm$0.1     &                 \\
$\Omega/h^-$                &               &               &               &               &0.0012$\pm$0.0005 &                 \\
$(\Omega+\overline\Omega)/h^-$ &            &               &               &               &0.0021$\pm$0.0008 &               \\
${\rm K}^0(892)/{\rm K}^-$   &               &               &               &              &0.26$\pm$0.08     & 0.23$\pm$0.05      \\
 \hline
$V$, $dV/dy$ ${\rm [fm}^3]$ &4871$\pm$394   &6082$\pm$384   &8204$\pm$351   &932$\pm$38      &930$\pm$3       &1182$\pm$55      \\
$T$ [MeV]                   &140$^*$        &141.9$\pm$0.5  &142.4$\pm$0.01 &142.2$\pm$0.01 &143.8$\pm$0.1    & 141.5$\pm$0.1     \\
$\lambda_q$                 &1.35$\pm$0.02  &1.25$\pm$0.01  &1.20$\pm$0.01  &1.15$\pm$0.02  &1.076$\pm$0.001  &1.062$\pm$0.001\\
$\lambda_s$                 &1.104$^*$      &1.074$^*$      &1.069$^*$      &1.054$^*$      &1.025$^*$        & 1.024$^*$     \\
$\gamma_q$                  &1.62$^*$       &1.62$^*$       &1.62$^*$       &1.62$^*$       &1.59$\pm$0.001   & 1.56$\pm$0.01     \\
$\gamma_s$                  &2.18$\pm$0.2   &2.20$^*$       &2.00$\pm$0.29  &2.13$\pm$0.14  & 2.22$\pm$0.01    &2.00$\pm$0.02      \\
$\lambda_{I3}$              &0.933$\pm$0.001&0.979$\pm$0.001&0.988$\pm$0.002&0.986$\pm$0.002&0.997$\pm$0.001  & 0.997$\pm$0.001 \\
 \hline 
$\mu_B$ [MeV]               &126            &94.8           &79             &61.2           &31.5             &25.7       \\
$\mu_S$ [MeV]               &27.7           &21.4           &16.5           &13.6             &7.0             &5.2      \\
 \hline
\end{tabular}\vspace*{0.1cm}
\end{center}
 \end{table*}

The local rest frame particle  density, $\rho={dN}/{dV}$,
is thus related to the rapidity density  by:
\begin{equation}\label{dNdy}
\frac{dN}{dy}= \frac{dV}{dy}\rho.
\end{equation}
The SHM fits to particle densities ${dN}/{dy}$ thus produce as
the normalization factor the value   $dV/dy$.
The qualitative relation between  $dV/dy$ and $V$
 (rest-frame hadronization  volume)  must include the
maximum rapidity range $ 2 y_p$, where $y_p$ is the rapidity 
of the nuclei colliding head on,
\begin{equation}
V=k {dV\over dy} \times 2 y_p,
\end{equation}
where $k$ is a reaction energy 
dependent constant. The study of the total hadron yields  
at RHIC we present  suggests $k\simeq 0.4$--0.6.

Regarding the data source, and weak decay acceptance, 
we need to consider  case by case the experimental results,
since the relative importance of hyperon decays in the total
baryon yields is high. In particular we note:
\begin{itemize}
\item  For RHIC-130 $dN/dy$ fit (second column from 
right in the top section of table \ref{Rhic}):\\
The $\pi^\pm,\ {\rm K}^\pm,\ p$ and $\bar p$ 5\% centrality results are from 
PHENIX \cite{Adcox:2003nr}. We assume that the ${\rm K}_S$ decays
into pions are accepted at 70\% level, and  ${\rm K}_L$ at 40\% level.
Mesons (pions and kaons) from hyperon decays are accepted at 30\% level,
while nucleons   from hyperon decay are nearly fully accepted, both 90\% and 99\%
acceptances   are in essence indistinguishable. $\Sigma^\pm$ decays  are 
fully accepted.  We include in the 
fit an average of the STAR~\cite{Adler:2002uv} and PHENIX~\cite{Adcox:2002au} 
 $\Lambda$ and $\overline\Lambda$ yields
where we can asses the feed from $\Xi$ and $\overline\Xi$ in view of the 
STAR analysis~\cite{Adams:2003fy}, we accept 99\% of
$\Omega$ and $\overline\Omega$  decays into  $\Lambda$ and $\overline\Lambda$.\\
For the $\Xi$ and $\overline\Xi$ weak feed yield  corrections are immaterial.
However, we cannot directly use the yields as these are presented for the 
10\% most central reactions. We fit the weak decay corrected $\Xi/\Lambda$ and 
$\overline\Xi/\overline\Lambda$  ratios. In order to relate  this to the total particle
yields, we include also  $\Xi/h^-$ ($h^-$ = negatives) where we accept in STAR $h^-$ 
the  weak decay products according to the pattern: ${\rm K}_S$ decays
into pions are accepted at 90\% level, and  ${\rm K}_L$ at 30\% level,
pions and kaons from hyperon decays are accepted at 50\% level,
while nucleons   from hyperon decay are  accepted, at  99\% level. The 
same is assumed in the fit  of $(\Omega+\overline\Omega)/h^-$ also  measured by 
STAR~\cite{Adams:2003fy}.  We include in the fit the STAR resonance ratios, 
${\rm K}^0(892)/{\rm K}^-$~\cite{Adler:2002sw}  and $\phi/{\rm K}^-$~\cite{Adler:2002xv},
in both cases we include 50\% feed from $\Omega$ and $\overline\Omega$ decay into kaons, which
is immaterial for the result. 
\item  For RHIC-200: $dN/dy$ fit (last column on
right in the top section of table \ref{Rhic}):\\ 
The $\pi^\pm,\ {\rm K}^\pm,\ p$ and $\bar p$ 5\% centrality results are from 
PHENIX \cite{phenixyield}. We assume that  the ${\rm K}_S$ weak decays are accepted 
 at 70\% level, and  ${\rm K}_L$ at 40\% level.
Mesons (pions and kaons) from Hyperon decays are accepted at 30\% level,
while nucleons   from hyperon decay are nearly fully accepted, we included 
this at 90\% level in the reported fit.  $\Sigma^\pm$ decays  are 
fully accepted.    We take STAR resonance ratios, 
${\rm K}^0(892)/{\rm K}^-$~\cite{Zhang:2004rj,Markert:2004xx} 
 and $\phi/{\rm K}^-$~\cite{phiyld},
in both cases we include 50\% feed from $\Omega$ and $\overline\Omega$ decay into kaons. 
The method to study the 
yields of stable hadrons along with resonances follows the 
work on the impact parameter dependence at $\sqrt{s_{\rm NN}}$
= 200 GeV~\cite{RafRHIC}.\\    
We did not use yields of $\Lambda$ and $\overline\Lambda$ since without direct
measurement of $\Xi$ and $\overline\Xi$ it is hard to judge the weak decay 
contamination in the data. Furthermore, we preferred to study  the relative 
yields $p/\pi^+, \bar p/\pi^-$. In the fit presented, we assumed that
the pion feed from WD of hyperons is at 80\% level. The other WD characteristics
are as discussed just above. This slight change in data input 
and also the slight modification of the pattern of weak decay acceptance has,
in comparison to Ref.\,\cite{RafRHIC}, yielded a increase of the 
volume factor $dV/dN$  by 1.2 s.d., while other variations are within
0.5 s.d..
\end{itemize}

We can expect, in near future, particle multiplicity 
results  from RHIC obtained at $\sqrt{s_{\rm NN}}$=62.4 GeV.
We interpolate the central rapidity yields of strange hyperons 
$\Lambda,\ \overline\Lambda,\ \Xi$ and $  \overline\Xi$,    presented 
in \cite{Elia:2004mb}, to this energy. With these 4 inputs, two constraints,   
setting the $\gamma_q=1.62$,  $T=140$, we find 
 a good description of the interpolated data but with   a few degrees 
of freedom. We have four interpolated `data' points, two constraints --- strangeness
conservation and $Q/b$, thus 6 data points which are  fitted using four flexible
parameters, $T,dV/dN,  \lambda_q, $ and $\lambda_{I3}$.
 This set of parameters, then, yields our
prediction of central rapidity particle multiplicities, 
seen in table \ref{outputRHIC},  for   $\sqrt{s_{\rm NN}}$=62.4 GeV.

We   make   an effort to understand   also the 
recently finalized total multiplicities $N_{4\pi}$ of K$^\pm$ and 
$\pi^\pm$~\cite{brahms}  at  $\sqrt{s_{\rm NN}}=200$ GeV.  
 Additional qualitative constraint  is derived from 
total charge particle multiplicities~\cite{Back:2002wb}, however
this result is not used directly in the fit. 
With the  three constraints, four BRAHMS particle yields,  we have 7 data points, 
and also 7 SHM parameters. To be able to make a 
 fit with  at least one degree of freedom it is necessary to make 
some `natural' hypothesis. For this reason, we do not 
discuss the fit  quality of  $N_{4\pi}$ yields at RHIC 
but we  discuss the expected total particle yields, which we regard 
to be an experimentally motivated hadron yield  prediction. 

We choose to consider   $\gamma_q=1.62\simeq e^{m_\pi/2T}$,
which we find systematically at the RHIC energy scale.  Our `fit' to $N_{4\pi}$ data 
at  $\sqrt{s_{\rm NN}}=200$ GeV works,  but it must not be seen as a full fit, 
rather a consistency test of SHM. This is allowing a prediction to be made of
other   $N_{4\pi}$ we show in table \ref{outputRHIC}. This 
consideration is also yielding   a rapidity-averaged value of $T$ and of the 5
chemical parameters, as well as an estimate of the proper size $V$
of the hadronizing fireball.  
The value of $\mu_{\rm B}$, which varies as function 
of rapidity, following the highly variable baryon distribution~\cite{Bearden:2003hx}, 
is found at a  median value, seen at the bottom of  table \ref{Rhic}, on left 
for the $N_{4\pi}$ fits.

\begin{table*}[!t]
\caption{
\label{outputRHIC}
Output hadron multiplicity data for the RHIC energy range. 
See text for the meaning of predictions of $N_{4\pi}$ yields at 
62.4 and 130 GeV and of  $dN/dy$ at 62.4 GeV.
The input statistical parameters are seen in table \ref{Rhic}.
 $b=B-\overline{B}\equiv N_W$ for $4\pi$ results and 
$b=d(B-\overline{B})/dN$ for results at central rapidity.
  Additional significant digits are presented
for purposes of tests and verification. All yields are without the 
weak decay contributions.
}\vspace*{0.2cm}
\begin{center}
\begin{tabular}{|c| c c c | c c c |}
\hline
$\sqrt{s_{\rm NN}}$  [GeV] & 62.4 & 130  &200 & 62.4 & 130 & 200  \\
$E_{\rm eq}$[GeV]     &2075 & 9008  & 21321 &2075 & 9008  &  21321 \\
$\Delta y$ &$\pm 4.2$&$\pm 4.93$&$\pm 5.36$ &$\pm 4.2$  & $\pm 4.93$ & $\pm 5.36$ \\
\hline
 &  \multicolumn{3}{c}{  $N_{4\pi}$\ 5\% } & \multicolumn{3}{|c|}{ $dN/dy|_{y=0}$\ 5\% }\\
\hline
$b $                       &350.2 &350.2 &349.6 &33.48 &18.50 &14.8 \\
$\pi^+$                    &899   &1201  &1543  &183.8 &230.3 &239.8 \\
$\pi^-$                    &927   &1229  &1573  &186.7 &231.9 &241.0 \\
${\rm K}^+$                &230.9 &302.5 &291.9 &43.7  &47.9  &47.1 \\
${\rm K}^-$                &168.5 &238.4 &242.3 &37,6  &44.2  &44.2 \\
${\rm K}_{\rm S}$          &193.8 &261.0 &259.9 &39.4  &44.4  &44.2 \\
$\phi $                    &27.3  &34.6  &28.9  &5.74  &6.86  &6.18 \\
$p $                       &140.0 &157.6 &192.0 &19.34 &17.09 &16.34 \\
$\bar p. $                 &24.1  &42.9  &66.1  &8.37  &11.11 &11.44 \\
$\Lambda$                  &81.1  &97.4  &89.9  &12.3  &12.04  &10.7 \\
$\overline\Lambda$         &20.2  &35.1  &38.3  &6.36  &8.60  &8.02 \\
$\Xi^-$                    &12.9  &16.4  &11.6  &2.14  &2.30  &1.91 \\
$\overline\Xi^+$           &4.6   &7.79  &6.13  &1.32  &1.80  &1.53 \\
$\Omega$                   &1.94  &2.68  &1.45  &0.36  &0.44  &0.33 \\
$\overline\Omega$          &1.04  &1.74  &0.98  &0.27  &0.38  &0.29 \\
${\rm K}^0(892)  $         &48.7  &67.4  &68.1  &10.2  &11.5  &11.2 \\
$\Delta^{0} $              &27.6  &31.1  &38.1  &3.78  &3.32  &3.15 \\
$\Delta^{++} $             &26.2  &29.9  &36.9  &3.69  &3.30  &3.13 \\
$\Lambda(1520)$            &4.43  &6.4   &6.0   &0.81  &0.81  &0.70 \\
$\Sigma^+(1385)$           &9.80  &11.91 &11.19 &1.52  &1.50  & 1.33\\
$\Xi0(1530) $              &4.20  &5.46  &3.88  &0.71  &0.78  & 0.64\\
$\eta $                    &131.9 &179.5 &192.3 &27.2  &30.5  & 30.6\\
$\eta' $                   &10.8  &15.2  &14.64 &2.30  &2.64  & 2.51\\
$\rho^0 $                  &85.8  &117   &157   &18.1  &19.5  & 20.3\\
$\omega(782) $             &75.9  &104   &142.8 &16.2  &17.4  & 18.3\\
$f_0(980)$                 &6.51  &9.03  &12.96 &1.40  &2.02  & 1.58\\
\hline
$(s-\bar s)/(s+\bar s)$    & 0   &  0   &  0   &  0   &   0   &  0    \\
\hline
\end{tabular}\vspace{0.1cm}
\end{center}
 \end{table*}

We extend the consideration of the  $N_{4\pi}$ yields to the 
lower energies,  $\sqrt{s_{\rm NN}}=62.4$ and 130 GeV. This can be done 
assuming that there is no change in   physics 
between top SPS energy  and RHIC 200 GeV run. Thus,
the success of our particle yield prediction would be  a confirmation of
this hypothesis. Our procedure can be seen in detail on the left of table  \ref{Rhic}.
We fix the hadronization temperature at $T=140$ MeV, choose the value  
  $\gamma_q=1.62\simeq e^{m_\pi/2T}$, and interpolate the values of $\gamma_s$. We
do find the required  values of $\lambda_q$, $\lambda_{I3}$ and $V$   needed to assure the total 
baryon yield, fraction of charge $Q/b$ and one particle yield, which we choose to be 
the interpolated total $\pi^+$.
We use the observation that the $\pi^+$ yield from Brahms
connects, in a logarithmic plot, in a  nearly perfect straight line
with the SPS energy domain. This produces the $\pi^+$ interpolated values 
we introduced in table \ref{Rhic}.  
The SHM succeeds perfectly and allows us to offer predictions for the total 
particle yields presented in table \ref{outputRHIC}.

We present, in detail,  the resulting particle multiplicities in  table \ref{outputRHIC} for
RHIC. On left, we show the expected total  yields  and on right the central rapidity yields.
We recall that, among total yields, only at  200 GeV   a significant experimental input 
was available, thus the 62.4 and 130 GeV total yield results are  an educated    
  guess satisfying all constraints and  criteria of the SHM model. 
Similarly, the central rapidity region yields for 62.4 GeV is a prediction based
on interpolated yields, with inputs seen in  table  \ref{Rhic}. All results,
presented  in  table \ref{outputRHIC}, are obtained prior to WD.

\subsection{Energy dependent particle yields}\label{energy}

\begin{figure}[!tb]
\hspace*{-.50cm}\epsfig{width=9.6cm,figure=\pathnow   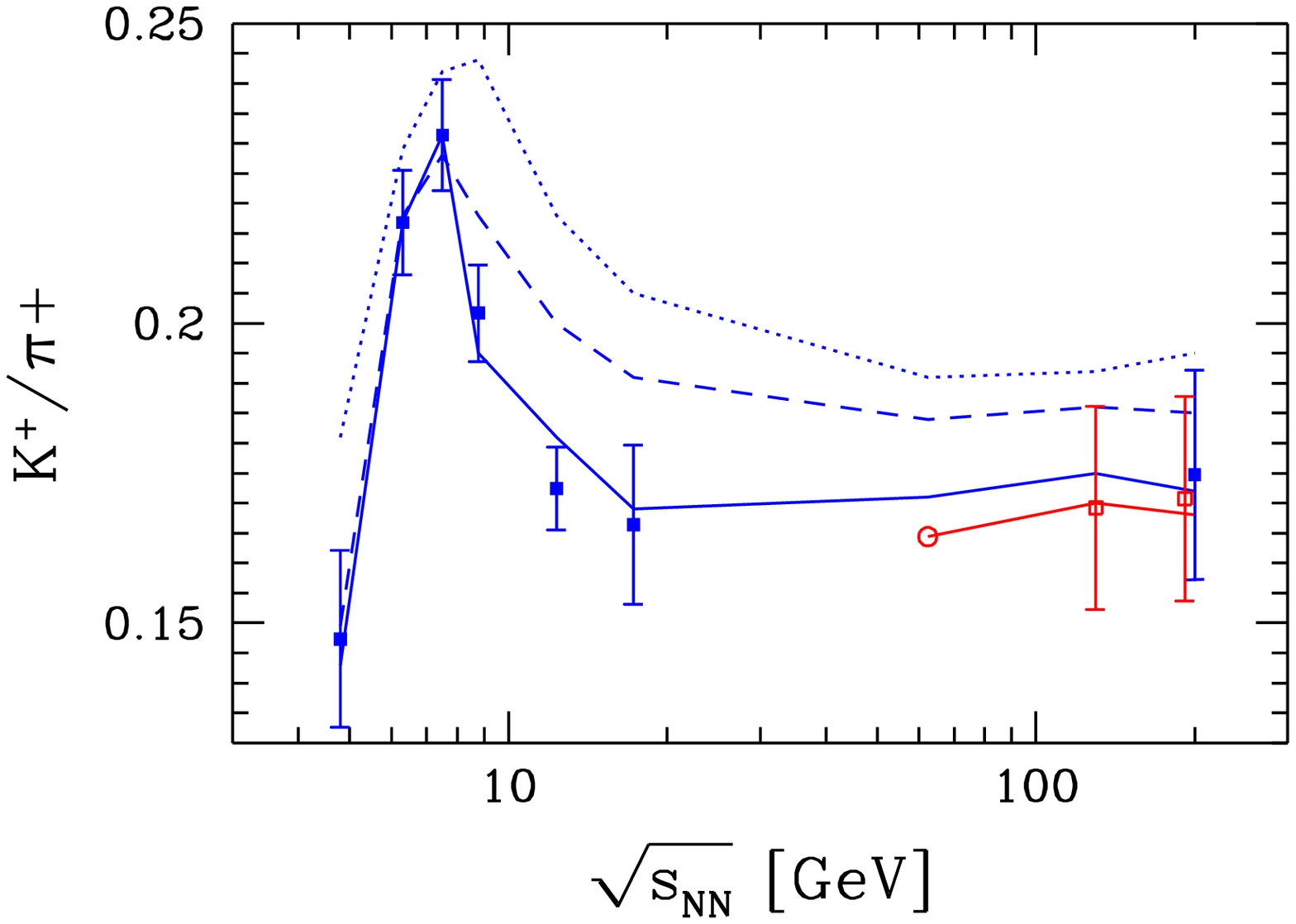}
\vskip -0.2cm
\caption{\label{KPi}
$K^+/\pi^+$ total yields  (filled squares, blue) and central rapidity density 
(open squares, red)
as function of  $\sqrt{s_{\rm NN}}$.
 The solid lines show chemical non-equilibrium  model fit.
 The chemical equilibrium
fit result is shown by the dotted line. The dashed line arises
finding best $\gamma_s$ for $\gamma_q=1$. See text about the total yield 
results at $\sqrt{s_{\rm NN}}=62.4$ and 130 GeV (unmarked edges in lines)
and about the central rapidity yield at  $\sqrt{s_{\rm NN}}=62.4$ (open circle).
}
\end{figure}

We consider, more systematically, the energy dependence of
particle yields and ratios.
Of particular interest is  the ratio  ${\rm K}^+/\pi^+$ 
which shows the previously unexplained horn structure. We
compare the experimental and theoretical behavior in  figure \ref{KPi}.
 The $4\pi$ results  are blue filled squares.  
 The central rapidity RHIC results (on right in red) are shown as open squares,
while the predicted total yield ratio for $\sqrt{s_{\rm NN}}=62.4$ is given
as an open circle.

We recall that the abrupt  increase in the  value of   $\gamma_q$ 
occurs    where    the rise in ${\rm K}^+/\pi^+$ reverses,  
turning into a sudden decrease with reaction energy.
 The solid line, shows our chemical non-equilibrium 
fit which   reproduces the horn structure  well. The
predicted total yield ratios   for $\sqrt{s_{\rm NN}}=62.4$ and 130 GeV (edges in solid line)
arise from the interpolation of yields and/or continuity in value of  statistical
parameters such as $\gamma_q$ between the top SPS and the top RHIC energy, see above, 
subsection \ref{RHfit}.

\begin{figure}[!tb]
\hspace*{-.40cm}\epsfig{width=9.5cm,figure=\pathnow  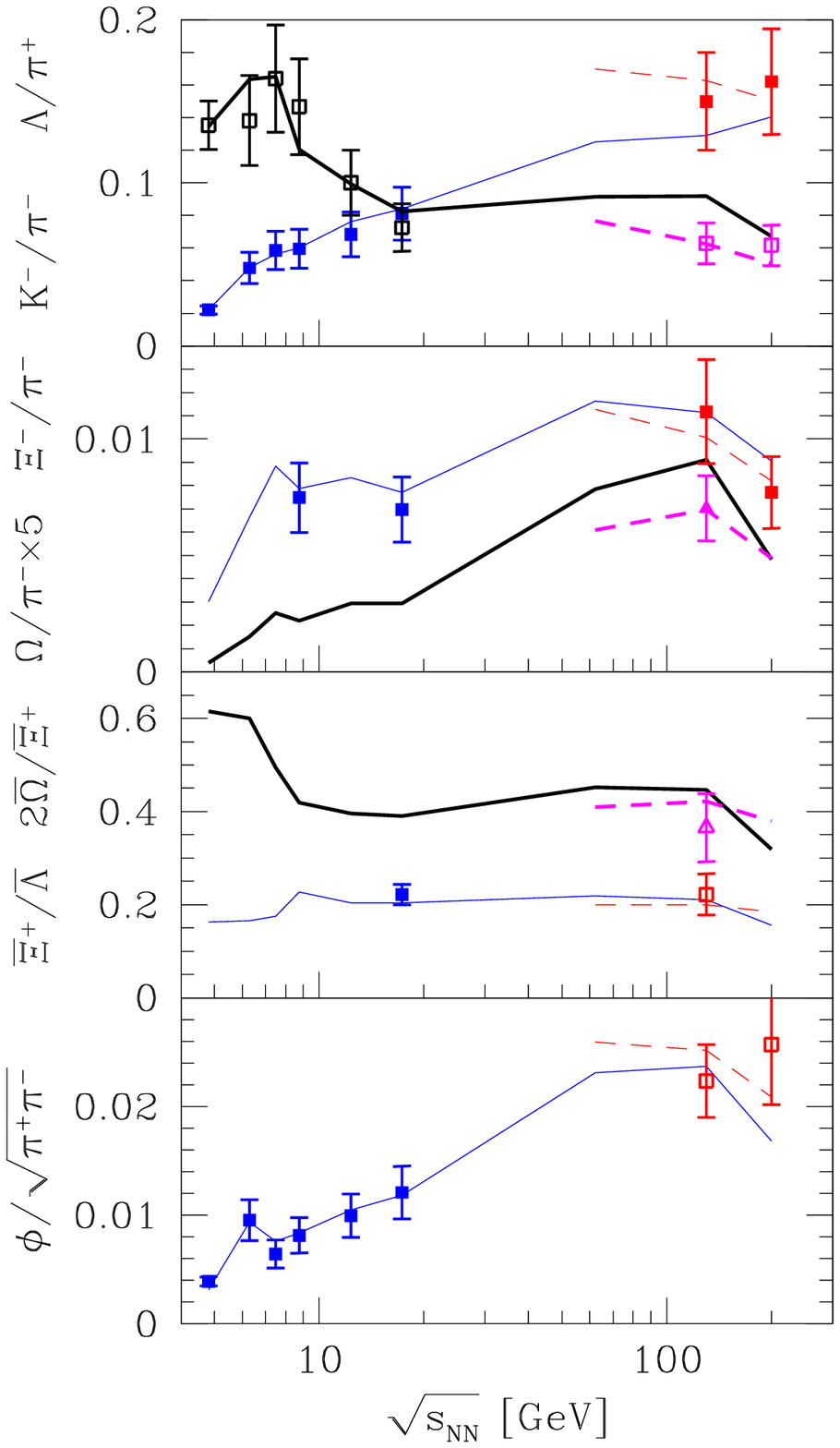}
 \caption{\label{LamAl}
Comparison of experimental and theoretical ratios of   particle yields
 as function of  reaction energy  $\sqrt{s_{\rm NN}}$ --- theoretical SHM
total $N_{4\pi}$  results are connected by  solid lines, with thick/thin lines corresponding to 
the different particle ratios. The  (corresponding) dashed   lines connect the
central rapidity $dN/dy$ results at RHIC. Experimental data used as fit input is shown
with its error bar.  
}
\end{figure}

The dotted line, in figure  \ref{KPi}, presents 
best fit results obtained within  the chemical equilibrium
 model, i.e., with $\gamma_s=\gamma_q=1$, using 
the same computer program (SHARE), and the same data set. 
We see  that the chemical equilibrium SHM cannot 
 explain the horn in the  ${\rm K}^+/\pi^+$ ratio. 
The dashed line corresponds to the result obtained fixing $\gamma_q=1$
but allowing $\gamma_s$ to assume a best value. We see that, without
 $\gamma_q> 1$, it is difficult if not impossible to obtain the large
reduction of  ${\rm K}^+/\pi^+$ ratio with increasing energy.
These findings are in line with prior attempts to explain the horn-structure,
 see, e.g., figure 4 in \cite{Cleymans:2004hj}. We note that our semi-equilibrium results follow 
better the trend set by the experimental data, which is a consequence 
of   the relaxation of exact strangeness conservation requirement. It appears that the full
chemical non-equilibrium statistical hadronization model is required in
order to obtain satisfactory understanding of the energy dependence of the
${\rm K}^+/\pi^+$ ratio. 

A graphic comparison of the experimental input, and 
theoretical output particle yields  as function of
energy for several other particles is seen  in  figure  \ref{LamAl}.
We show  ${\rm K}^-/\pi^-$ with $ \Lambda/\pi^+$, $\Xi^-/\pi^-$ with 
$5\,\Omega^-/\pi^-$,
$\overline\Xi^+/\overline\Lambda$ with $2\,\overline\Omega/\overline\Xi^+$
and at bottom $\phi/\sqrt{\pi^+\pi^-}$.
We are showing  the total SHM yield ratios  at AGS/SPS as well as at  RHIC, connected
by a solid line (thick and/or thin).
The central rapidity yields at RHIC are also presented for comparison by the dashed lines.
 (in red and violet) . 

The SHM allowing for chemical non-equilibrium 
reproduces all salient features of the experimental
particle yield data well as function
of energy, including the NA49 results that otherwise 
could not be described in equilibrium and semi-equilibrium 
approach~\cite{Blume:2004ci}, e.g., the already discussed  
$K^+/\pi^+$ ratio shown in figure \ref{KPi}. In addition, In figure \ref{LamAl}, 
we note in the top panel the    shift of $s$ quark population
from its dominant baryon component (see $\Lambda/\pi^+$) at low reaction 
energy, to meson  carriers (see K$^-/\pi^-$).

\begin{figure}[!tb]
\hspace*{-.40cm}\epsfig{width=9.5cm,figure=\pathnow  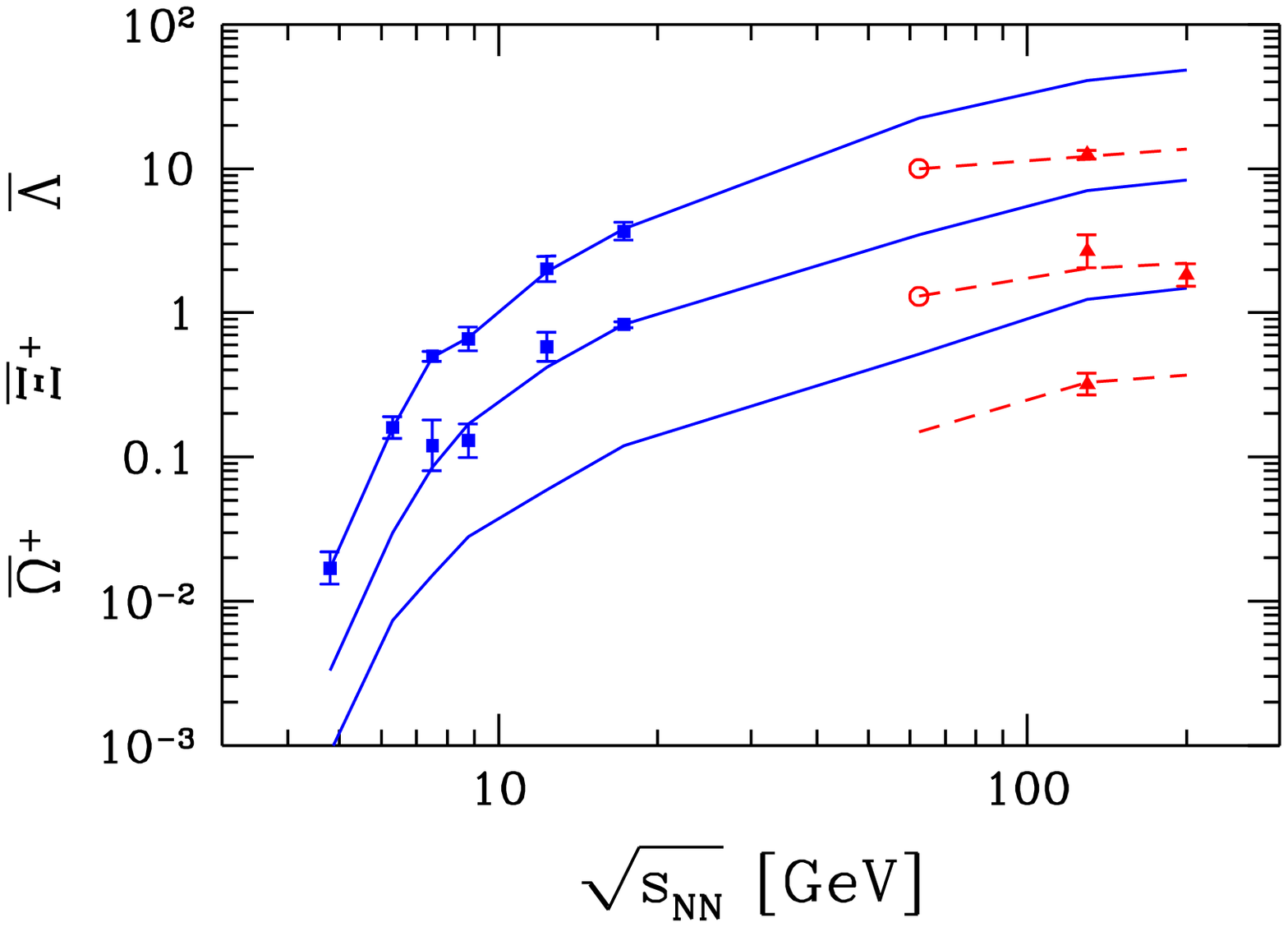}
 \caption{\label{Figantibaryon}
Yields of strange antibaryons as function  $\sqrt{s_{\rm NN}}$, from top to bottom 
$\overline\Lambda, \overline\Xi^+$ and $  \overline\Omega^+$.
The solid lines connect  the results of SHM  $N_{4\pi}$ fit to particle data. 
 The AGS/SPS energy range
 $N_{4\pi}$ yields (in blue)  on  left   and RHIC  $N_{4\pi}$ on right. Also 
on right (in red, connected with dashed lines) are the central   $dN/dy$ yields.   
 The   yields  at  $\sqrt{s_{\rm NN}}=62.4$ used in our study are result of
interpolation of RHIC and SPS results.  
}   
\end{figure}

Of particular importance, in the study of quark--gluon plasma
formation, is strange antibaryon enhancement. It is one
of important signatures of deconfinement~\cite{Rafelski:1982ii}. 
These particles are hard to make in conventional environment, and
also are highly sensitive probes of the medium from which they 
emerge. There is still  only fragmentary   
data available for antibaryon ratios of interest,
$\overline\Xi^+\!\!/\overline\Lambda$ and  $2\overline\Omega/\overline\Xi^+\!\!,$ 
shown in figure  \ref{LamAl}, in the third panel from the top.

In figure  \ref{Figantibaryon}, we show   as function of energy   the individual 
yields of $\overline\Lambda,\ \overline\Xi^+$ and $  \overline\Omega^+$, 
along with the  experimental data  used as input.  All three antihyperon production 
rates are predicted to rise at nearly the same rate
 as function of  $\sqrt{s_{\rm NN}}$ --- indeed the 
strange antibaryon ratios, we have shown in figure \ref{LamAl},
are as good as  flat compared to the great variability of the   absolute 
yields, which are  increasing very rapidly, as is seen
 in figure  \ref{Figantibaryon}.    The  strange antibaryon production
enhancement phenomenon has been   considered in terms of a comparison of 
  yields of antibaryons to a baseline
yield obtained scaling the $pp$ or $p$Be yields. This was done   
as function of impact parameter, and reaction energy~\cite{Bruno:2004pv}.

We note that the SHM non-equilibrium approach   under-predicts by  1.5 s.d. 
the yields of $\Omega$ and $\overline\Omega$, for both SPS-NA49 results available
at 40 and 158 $A$ GeV. We did not include 158 $A$ GeV results
in our input data set since non-SHM source, 
such as  chiral condensate~\cite{Kapusta:2000ny}, can generate such
an excess,

Ratios of strange antibaryons are a 
sensitive probes of the medium.
Once the deconfined phase is formed, the ratios
of yields of strange antibaryons should not change drastically.
 Thus, our  finding as function of energy   
  in essence of a  flat ratio, with minor fluctuations 
originating in the other experimental data and amplified by the 
sensitivity of these particles, suggest that the same form of (deconfined) matter is
present at SPS and RHIC, except perhaps for the lowest SPS reaction energy.

Another   important point is that these antihyperon ratios are relatively large, and
hard to understand except  in terms of the 
quark  coalescence picture. It would be very 
interesting to  confirm experimentally that,  at AGS energy scale,  $2\overline\Omega/\overline\Xi^+$ is
indeed as large as predicted in figure  \ref{LamAl}. This would 
establish coalescence quark chemistry  in this low energy  environment. Further, 
this maybe taken as an indication
that the transition we observer at  $\sqrt{s_{\rm NN}^{\rm cr}}$
involves two deconfined phases of different structure. We will further 
discuss this in  section~\ref{results}.

\subsection{Yields of pentaquark hadrons}\label{pentas}
There are now more than 600 papers with the title word `pentaquar', however,
on balance the evidence for the  exotic hadrons $\Theta^+(1540)$ with the 
quark content $[uudd\bar s]$, and typical decays $p{\rm K}^0, nK^{*+}$, and  
$\Xi^{-\,-}(1862) [ssdd\bar u]$, with typical decays  $\Xi^-\pi^-, \Sigma^-{\rm K}^- $,
is not   convincing. The $\Xi^{--}(1862)$  may
have been observed by NA49 in $pp$ interactions 
at top SPS energy~\cite{NA49XIPQ}. The $\Theta^+(1540)$
remains uncertain: several high statistics confirmation experiments  failed to find
this state. Arguments were presented why these states were incorrectly identified~\cite{Dzierba:2003cm}. 
We have therefore not included these and other related
exotic quark states from the hadron resonance list in SHARE, when performing the 
fits here presented.

\begin{table*}[!t]
\caption{
\label{outputPenta}
Predicted yields of  $\Theta^+(1540)$ and $\Xi^{--}(1820)$ pentaquarks    for AGS, SPS and RHIC, 
obtained with  SHM parameters shown in  tables \ref{AGSSPS} and \ref{Rhic}.  
}
\vspace*{0.2cm}
\begin{center}
\begin{tabular}{|c| c | c c c c c c c c | }
\hline
$E$ [$A$\,GeV]             &11.6 & 20  & 30    & 40    & 80  & 158   &2075  & 9008  & 21321 \\
$\sqrt{s_{\rm NN}}$  [GeV] &4.84 &6.26 &7.61   &8.76   &12.32 &17.27 & 62.4 & 130  &200 \\
\hline $N_{4\pi}$ centrality      &m.c. &  7\% &  7\% &  7\% &  7\% &  5\%&  5\%&  5\%&  5\%  \\
$\Theta^+(1540)$           & 0.66 &1.14 & 6.90 &7.15  &6.52  &6.70  &7.23  &7.92  &7.19 \\
$\Xi^{--}(1820)$           &0.0022&0.010& 0.098&0.11  &0.18  &0.24  &0.71  &0.89  &0.78  \\
\hline
\end{tabular}\vspace*{0.1cm}
\end{center}
 \end{table*}
  
On the other hand,   theoretical arguments for the existence of pentaquark states
have not been refuted. We thus   present in  table  \ref{outputPenta} 
 predictions for the production rates of 
$\Theta^+$   and  
$\Xi^{--}$. These yields are highly sensitive to the hadronization
conditions~\cite{Letessier:2003by}, and were obtained 
using the parameters of the fits here presented.

 The expected statistical hadronization
yield of $\Theta^+$
 rises rapidly, by an order of magnitude, between 11 and  30 $A$\,GeV 
reaction energies and remains practically constant thereafter.  
The expected $\Theta^+(1540)$ yield  in fact
exceeds the SHM predicted yield 
of $\Lambda(1520)$ in the threshold energy domain by a factor 2--4,
and comparing to the observed $\Lambda(1520)$  yield at 158 $A$ GeV by more
than a factor 4. Furthermore, at 30 $A$ GeV   
the background multiplicity is relatively small, 
while the rapidity range is also restricted
compared to the top SPS energy, which should help finding the 
pentaquark, if it exists 
in the range of energies characterizing the horn in the K$^+/\pi^+$. 

\section{Fireball properties at breakup}\label{Fire}
\subsection{Energy dependence of  model parameters} \label{statpar}
\begin{figure}[!bt]
\hspace*{-.60cm}\epsfig{width=9.9cm,figure=\pathnow    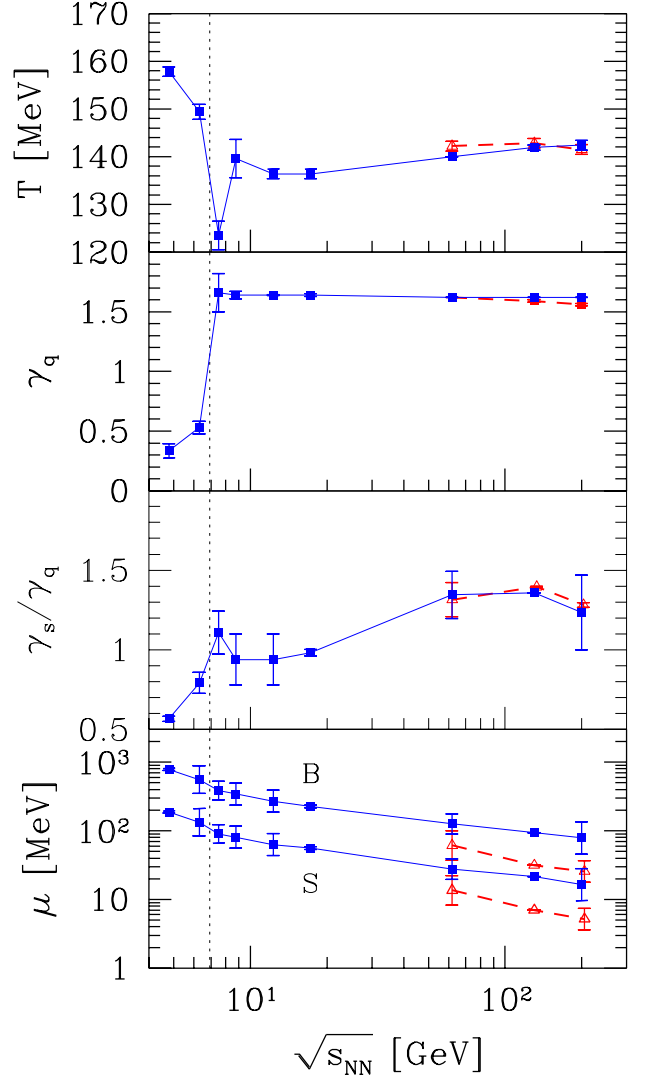}
\caption{\label{gammu}
 Statistical parameter results for $N_{4\pi}$ (blue online, square). From  
 top to bottom: $T$, $\gamma_q$, $\gamma_q/\gamma_s$ and
 $\mu_{\rm B}, \mu_{\rm S} $[MeV],
as function of  $\sqrt{s_{\rm NN}}$. The lines guide the eye. 
 Same   for $dN/dy$ at RHIC (red online triangles).
}
\end{figure}

The   statistical parameters of the SHM   are shown, as function of 
$\sqrt{s_{\rm NN}}$, in figure \ref{gammu}, for the entire energy domain. From
 top to bottom, we see the chemical freeze-out temperature $T$, 
the statistical occupancy parameters $\gamma_q$ and  $\gamma_s/\gamma_q$
and the chemical potentials  $\mu_{\rm B}$ and $\mu_{\rm S}$.
The error bars
comprise the propagation of the experimental yield errors, as
well as any uncertainty due to the shape of the 
$\chi^2/{\rm d.o.f.}$ minimum,   seen  in figure \ref{ChiP}. 
The (red) triangle results are for the RHIC $dN/dy$ case, while 
(blue) squares are for the $N_{4\pi}$ data throughout the energy
domain and include the estimates we made for the RHIC energy range
(at $\sqrt{s_{\rm NN}}=62.4$ and 130 GeV, we do not show for these 
 fits an  error bar, as these results are  solely our  estimate).

The only significant difference between RHIC $dN/dy$ and $N_{4\pi}$
results is noted for the chemical potentials $\mu_{\rm B}$ and $\mu_{\rm S}$ and
shown in the bottom panel (note logarithmic scale).  
The baryochemical potential  $\mu_{\rm B}$
drops relatively smoothly as the reaction energy is increased. The
vertical line indicates the observed sudden change in the structure 
of the fireball. This is seen in all statistical variables,
but most clearly in $\gamma_q$.

It is important to recall that we present $\gamma_i$ evaluated
using hadronic multiplicities. If these arise from breakup
of a quark fireball, the quark-side occupancy
parameters could be considerably different. The hadron-side phase
space size is, in general, different from the quark-side phase space,
since the particle degeneracies, and  masses,  are quite different. 
In the study of the  breakup of the quark fireball into hadrons, we can 
compute the resultant hadron phase space occupancy for two 
extreme limits.  

First, consider  a fast transformation (sudden breakup)
 of the quark phase. This occurs nearly  at fixed volume.
To accommodate the difference in the momentum part 
of the phase space, the   chemical occupancy 
non-equilibrium parameters $\gamma_i$  undergo 
an abrupt change. We note that 
 it is of no importance if there was
or not a phase transition between the phases, what 
matters is that there was no time to reequilibrate 
chemically the quark yields.  
In the opposite limit of a very slow transformation
of phases, there is available a prolonged period in time in which 
the volume of the system can change to accommodate  
the appropriate number of particles in chemical equilibrium
corresponding to the maximum  entropy content.

To determine the change in $\gamma_i$ in sudden hadronization, one
needs to compare in detail the phase space of quark phase with that
of hadron gas.  In order
to make this comparison  one must consider the energy and entropy
content of the QGP phase.  
For $\mu_{\rm B}=0$, as well as small chemical potentials  
$\mu_{\rm B}/T< 1$,  lattice evaluation of
the deconfined phase properties are available 
\cite{Aoki:2006we,Aoki:2006br,Fodor:2004nz,Allton:2002zi}.
 It is thus possible to model quantitatively the 
properties of the deconfined phase, and to compare these with the 
results of   the SHM~\cite{Kuznetsova:2006bh}.
The remarkable result   is that near to $T=140$ MeV the 
sudden transition requires  the value $\gamma_q\simeq 1.6$ on the 
hadron side, if the quark phase was chemically equilibrated.

 Thus, the large values of $\gamma_s^{\rm HG}$ and $\gamma_q^{\rm HG}$,
seen in  figure \ref{gammu} at large $\sqrt{s_{\rm NN}}$,
where $\mu_{\rm B}/T< 1$
are  consistent with sudden breakup of chemically equilibrated  
primordial QGP phase. Other dynamic effects, in particular fast
expansion, in general also favor an over-saturated phase space with $\gamma_i> 1$. 

As seen in figure \ref{gammu}, 
$\gamma_s/\gamma_q$ rises at first rapidly, as expected
if strangeness production is delayed by  a greater 
threshold mass and has to catch up with the light hadron production. 
$\gamma_s/\gamma_q$  decreases beyond the edge of energy threshold,
as can be expected due to the conversion  of the quark to hadron occupancy 
discussed above. The rise resumes and continues for all energies
above 80  $A$GeV (note that,    at RHIC only,    results showing an 
error can be considered to arise from  a fit).

\subsection{Physical properties}  \label{physres}
We now turn our attention to the physical properties of the hadronizing 
fireball obtained summing individual properties of   hadronic particles  
 produced. One can view the consideration of the physical properties
of the fireball at breakup as another way to present the SHM parameters. 
For example, the net baryon density, $\rho_{\rm b}\equiv (B-\overline B)/V$,
is most directly related to the baryochemical potential $\mu_{\rm B}$,
the thermal energy density $E_{\rm th}/V$ is related to $T$ etc. 
 
We present the  physical properties, i.e., pressure $P$,  energy density $\epsilon$,
entropy density $S/V$, net baryon density $\rho_{\rm b}\equiv (B-\overline B)/V$ and
the yield of strangeness $s$,  in  table \ref{AGSSPSPhysical} for the AGS/SPS 
energy range considered. Note that $s$  contains 
hidden strangeness from $\eta,\ \phi$ and $\eta'$. At the bottom of table  
\ref{AGSSPSPhysical}, we show 
the dimensionless ratios of extensive variables 
$P/\epsilon$, and $E_{\rm th}/TS$. These two  ratios are  very smooth as
function of energy, and lack any large fluctuations that could be
associated with fit error. These ratios are characteristic for the   
conditions of the fireball at the point of hadronization. 

The results presented can be used to constrain 
dynamical models describing the evolution of the QGP fireball in time towards
hadronization/particle freeze-out. We present the energy range at RHIC on left
in table \ref{Rhicphys}. We recall that  the  62.4 GeV
 and the 130 GeV $4\pi$ results, as well as in part the 200  $4\pi$ results,
 are  result of considerations which do not involve experimental measured
particle yields. Thus, the $4\pi$  results are to be seen as SHM
sophisticated prediction. On right, in table  \ref{Rhicphys}, we
present the results for central rapidity densities. Here, only 
the   62.4 GeV case is a prediction, the other results are direct
consequence of the data interpretation in terms of SHM.

\begin{table*}[!t]
\caption{
\label{AGSSPSPhysical}
The physical properties: Pressure $P$, energy density $\epsilon=E_{\rm th}/V$, entropy density $S/V$,
strangeness density $s/V$  for AGS and CERN energy range at,  
(top line) projectile energy $E$ [GeV].  Bottom: dimensionless ratios of properties at fireball breakup,
$P/\epsilon$ and $E_{\rm th}/TS$ .
}\vspace*{0.2cm}
\begin{center}
\begin{tabular}{|c| c | c c c c c |  }
\hline
E[$A$GeV]                   & 11.6        & 20          & 30          & 40         & 80            & 158  \\
$\sqrt{s_{\rm NN}}$  [GeV]  &4.84         &6.26         &7.61         &8.76        &12.32          &17.27  \\
\hline\hline
$P{\rm [MeV/fm}^3]$         &21.9         &21.3         &58.4         &68.0         &82.3         & 76.9            \\
$\epsilon{\rm [MeV/fm}^3]$  &190.1        &166.3        &429.7        &480.2        &549.9        & 491.8            \\
$S/V{\rm [1/fm}^3]$         &1.25         &1.21         &2.74         &3.07         &3.54         & 3.26            \\
100$\bar s/V {\rm [1/fm}^3]$&0.988        &1.52         &5.32         &5.85         &7.65         & 7.24      \\
$\rho_b{\rm [1/fm}^3] $     &0.104        &0.0753       &0.184        &0.186        &0.167        & 0.121            \\
   \hline
 $P/\epsilon$               &0.115        &0.128        &0.136        &0.142        &0.150        & 0.156          \\
$E_{\rm th}/TS$             &0.96         &0.92         &1.27         &1.20         &1.14         & 1.11         \\
   \hline
\end{tabular}\vspace*{0.1cm}
 \end{center}
\end{table*}

\begin{table*}[!t]
\caption{
\label{Rhicphys}
 The physical properties for RHIC energy 
range,  see table \ref{AGSSPSPhysical} for details.  For the central rapidity 
case,  we show the rapidity densities: energy  rapidity density 
$d\epsilon=dE_{\rm th}/dV$, entropy  rapidity density $dS/dV$,
strangeness rapidity  density $ds/dV$ and net baryon rapidity density  $db/dV$.  
All  62.4 GeV results, and the 130 GeV $4\pi$ result
 are,  as discussed in text, result of assumptions, and/or interpolations of 
yields and/or parameters, and hence are a  prediction.
}\vspace*{0.2cm}
\begin{center}
\begin{tabular}{|c| c  c c | c c c |  }
\hline
$\sqrt{s_{\rm NN}}$  [GeV]   & 62.4      & 130     &200    & 62.4 & 130 & 200  \\
\hline
 &  \multicolumn{3}{c}{  $N_{4\pi}$ } & \multicolumn{3}{|c|}{ $dN/dy|_{y=0}$ }\\
 \hline
$P  {\rm [MeV/fm}^3]$       &82.4        &87.8     &80.0   &80.5  &91.4  & 94.5       \\
$dE_{\rm th}/dV {\rm [MeV/fm}^3]$ &516.6 &548.4    &478.9  &532.5 &604.4 & 479.4      \\
$dS/dV {\rm [1/fm}^3]$      &3.62        &3.73     &3.32   &3.64  &4.03  & 3.32       \\
100$d\bar s/dV {\rm [1/fm}^3]$ &11.5     & 12.4    &9.2   &12.0  &13.7  & 10.4      \\
100$db/dV {\rm [1/fm}^3] $  &7.19        & 5.76    &4.26   &3.59  &1.99 & 1.26     \\
 \hline
$PdV/dE_{\rm th}$           &0.159       & 0.160   &0.167  &0.151 &0.151 & 0.197       \\
$dE_{\rm th}/TdS$           &1.02        &1.04     &1.01   &1.03  &1.04  & 1.02        \\
 \hline
\end{tabular}\vspace*{0.1cm}
 \end{center}
\end{table*}

The fit uncertainty in the  quantities
presented in tables \ref{AGSSPSPhysical} and \ref{Rhicphys}
 is difficult to evaluate  in detail. The individual
physical properties require powers and exponents 
of statistical parameters, and thus, at first sight, 
we expect that   the fractional errors are increased,
as compared to those prevailing among 
statistical parameters in    table \ref{AGSSPS}.
However,  the dominant contributions to each physical
 property  is often  directly 
derived from the individual  observed particle yields.
Therefore,  a large  
compensation of  errors originating in the fitted statistical  
parameter errors must occur.  

\begin{figure*}[!bht]
\hspace*{-.5cm}\epsfig{width=9cm,angle=0,figure=\pathnow 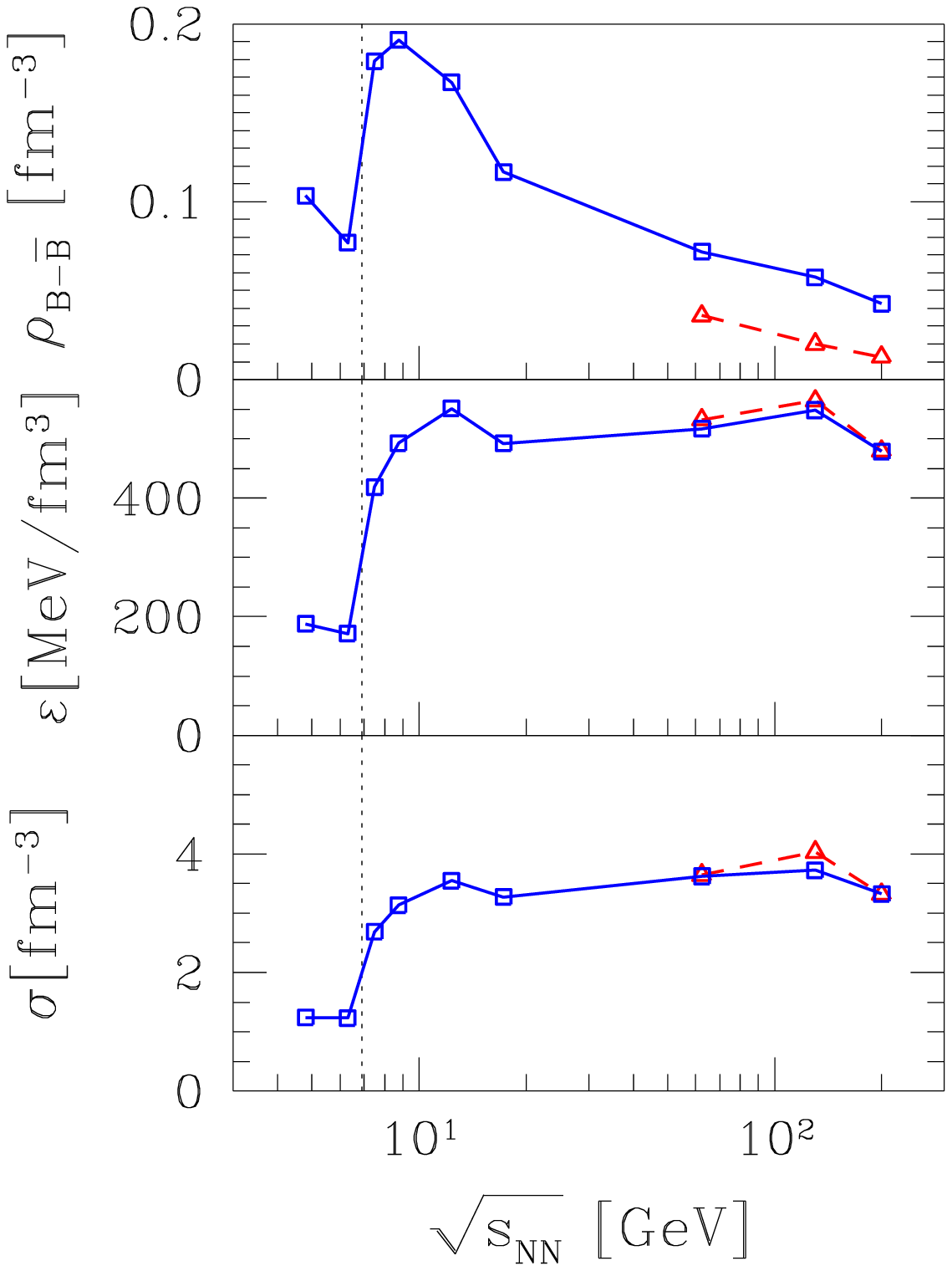}
\epsfig{width=9cm,angle=0,figure=\pathnow 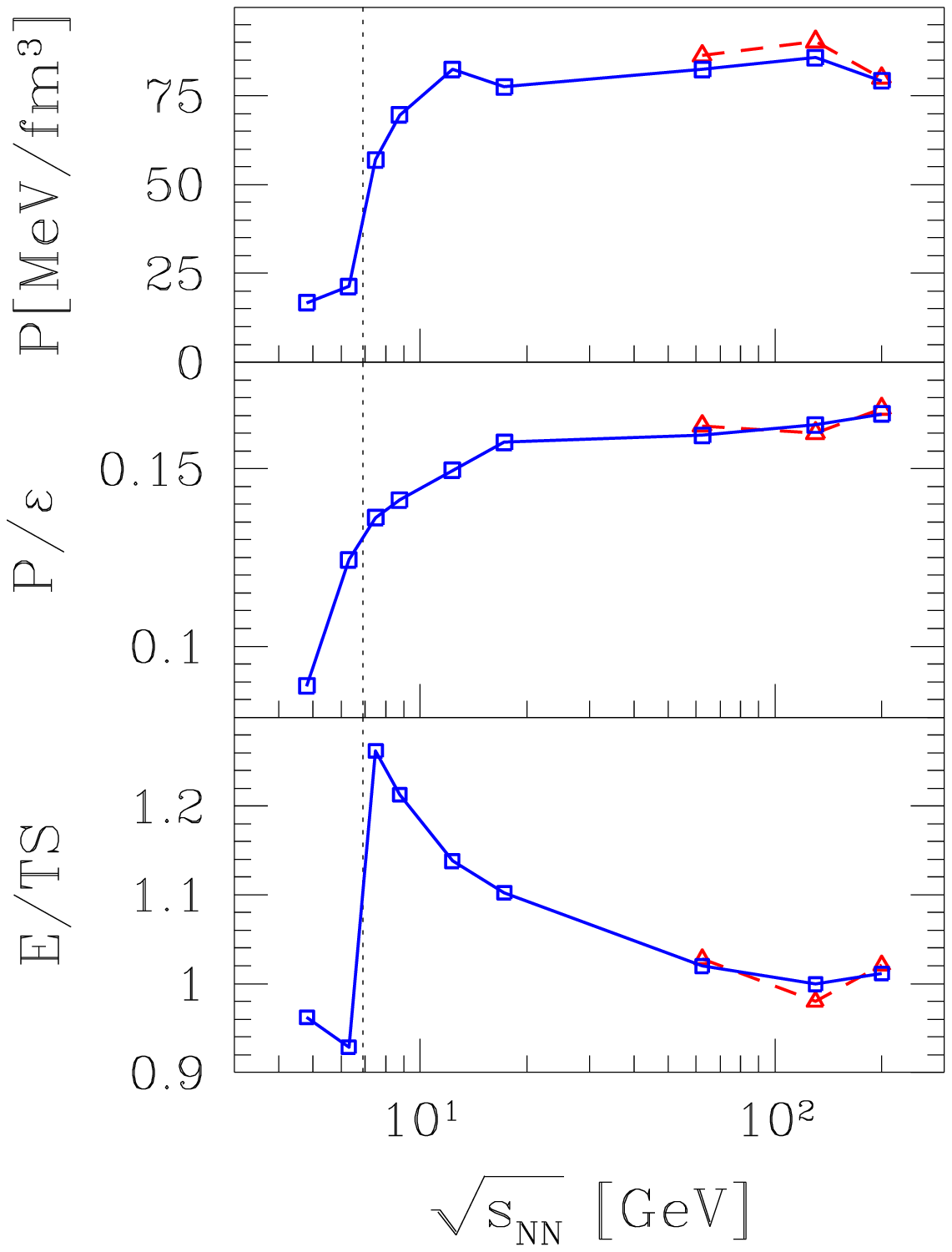}
\caption{\label{physprop}    From 
 top to bottom, on left hand side the baryon density 
$\rho_{B-\overline B} [{\rm fm}^{-3}]$, energy density $\epsilon[{\rm MeV}/{\rm fm}^3]$, and 
entropy density $\sigma[{\rm fm}^{-3}]$, as function of 
 $\sqrt{s_{\rm NN}}$, on right
hand side pressure $P[\ {\rm MeV}/{\rm fm}^{3}]$,
 $P/\epsilon$, and $E_{\rm th}/TS=\epsilon/T\sigma$. 
Squares (blue) average over the entire fireball 
at hadron freeze-out,   triangles (red)
for the   central rapidity region of the fireball.
}
\end{figure*}

For example,  most of the pressure at breakup is 
due to the most mobile, lightest particle, the pion. These yields are 
known to better than 10\%, and thus, the pressure must be known to greater 
precision since there are further constraints from consistency of this
yield with the yield of other particles. This explains why the 
results when presented graphically  (see figure \ref{physprop})
are at 5\% level smooth functions of $\sqrt{s_{\rm NN}}$, with
fluctuations  apparently at worse similar to  those we see in the
individual  statistical SHM parameters. In  future, one could hope to 
fit to the experimental data directly the physical properties, bypassing
the statistical parameters.  This can be done, in principle, considering the 
mathematical properties of these expressions. However,
such study transcends considerably the scope of this paper, and it is
indeed  motivated by results we present for the first time here. 

On the left hand side, in  figure \ref{physprop}, we see
from top to bottom the baryon density, the
thermal energy density $\epsilon$ and the entropy density $\sigma$. 
On the right hand side, from top to bottom, 
we show the pressure $P$, and the dimensionless ratio  
of pressure to thermal energy density  $P/\epsilon$,
and $E_{\rm th}/TS=\epsilon/T\sigma$. 
The triangles (red) correspond to the properties of the fireball  
at central rapidity at RHIC energy scale. We note a significant 
 difference  between the total
fireball averages (squares) and the   central rapidity result (triangles)
only  in the net baryon number density.

As the reaction energy passes the threshold,  
$6.26\,{\rm GeV}$ $<\sqrt{s_{\rm NN}^{\rm cr}}<7.61\,{\rm GeV}$,
the hadronizing fireball 
becomes much denser: the   entropy density jumps by factor 4, 
and the energy and baryon number density by a factor 2--3.
The hadron pressure increases from $P=25\ {\rm MeV}/{\rm fm}^3$ initially
by factor 2, and ultimately more than factor 3. There is a more
 gradual increase of $P/\epsilon=0.115$ at low reaction energy 
to 0.165 at the top available energy. Also $E_{\rm th}/TS$ falls gradually
from 0.9 down to  0.78  for   the high density fireball. 

The rather rapid change in the individual  properties: entropy,  energy,  pressure
is  seen, in figure \ref{physprop}, to be largely compensatory, resulting 
in a smooth change in $P/\epsilon$, and similarly  $E_{\rm th}/TS$. 
Even though  there is 
a small residual variation reminding us of the sudden changes in 
the three factors  in the ratio  
$E/TS$, this quantity is extraordinarily  smooth. Moreover, we see
the same value for the central rapidity as we find for the average over
the entire fireball. Thus $E/TS\simeq 0.78$  could be a universal 
hadronization constraint.  

For AGS 11 GeV and SPS 20 GeV results, the value $E/TS$ is
greater, reaching to  $E/TS\simeq 0.9$. This requires $P/\epsilon$ 
to be smaller, as is the case when the effective quark mass increases. 
A simple structure model employing a thermal quark mass, $m_q\simeq aT$,
was considered in  Ref.~\cite{RafRHIC}. $E/TS\simeq 0.78$ corresponds to
the value of $a\simeq 2$ (usual for thermal QCD) and found in the limit
of  large $A$ and here large $\sqrt{s_{\rm NN}}$. $a$ has to rise
to $a \ge 4$ in order to explain the rise in $E/TS$. This points
to a phase of `heavy' quarks being at the origin of the increase of $E/TS$ 
with decreasing  $\sqrt{s_{\rm NN}}$. Such a heavy quark phase can be possibly
a `valon' quark phase and we pursue this further in section \ref{bound}.

We believe that any structure model 
of the   phase transformation, and/or the two phase structure will need to 
address  $E/TS$, and/or $P/\epsilon$ freeze-out condition results quantitatively.
These two ratios, $E_{\rm th}/TS$ and $P/\epsilon$,
are  related.  Restating the 1st law of thermodynamics: 
\begin{equation}\label{1stlaw}
\frac{E_{\rm th}}{TS}(1+k)=1+\frac{\Sigma_i\nu_i \ln\Upsilon_i}{\sigma}, 
 \quad k={P\over \epsilon} .
\end{equation}
For each hadron fraction with density $\nu_i$ the total fugacity is 
\begin{equation}
\Upsilon_i=\prod_j\gamma_j^{n_j}\lambda_j^{n_j}, 
\end{equation}
where all valance quarks and antiquarks of each hadron fraction
contribute in the product,  see section 2 in \cite{share}. In
the limit of chemical equilibrium:
 \begin{equation}
\frac{\Sigma_i\nu_i \ln\Upsilon_i}{\sigma}\to 
   \frac{ \rho_{B-\bar B}\mu_{\rm B}}{T\sigma}.
\end{equation}
Thus, in this limit  at the RHIC energy range, we expect that 
$E_{\rm th}/TS\to 1/(1+P/\epsilon)$. However, the results in figure \ref{physprop}
show, the chemical non-equilibrium effects contribute considerably.

It is interesting to note that  the same behavior of the 
physical properties of the fireball  has  also been obtained 
 as function of the  volume in the study 
of impact parameter dependence, see figure 4 in \cite{RafRHIC}.
In fact, the results we derived  show an unexpected universality of the 
hadronizing fireball, which depends solely  on the question if it
occurs `below' or `above' the threshold in energy and volume size;
the volume threshold corresponds to critical participant 
number  $13.4< A^{\rm cr}<  25.7$. 

At these values of $A$ and the associated baryon content at
central rapidity~\cite{RafRHIC}, the grand canonical description of particle
yields is still justified~\cite{Rafelski:1980gk,Rafelski:2001bu}, 
also for strangeness. However, the fitted reaction volume (not further used 
in the present work)  may be revised within the canonical approach
by 10 -- 20\% for the most peripheral collisions 
studied in~\cite{RafRHIC}. 
 
\subsection{Strangeness and Entropy yield}\label{SE}
 The yield
 of strangeness produced, should the  deconfined QGP fireball 
be formed, is sensitive to the initial conditions,
especially to the temperature achieved. 
 The   standard results for strangeness  relaxation time corresponds to 
$\tau_s(T=300\,{\rm MeV})\simeq 2$fm/c \cite{Letessier:1996ad}.  When this result is
used in model calculations addressing RHIC~\cite{Rafelski:1999gq},
one finds, assuming gluon thermal and chemical equilibrium, 
that the thermal strangeness production in the early stage suffices to 
  saturate  the QGP fireball phase space at hadronization.

Even so, there is considerable uncertainty how short the time required to
relax strangeness flavor  is, as the
relaxation time lengthens with the square of the  glue phase space
under-occupancy, $\tau_s\propto 1/\gamma_{\rm G}^2$.  
Much  of the uncertainty 
about the gluon chemical conditions prevailing in the 
initial thermal phase can be eliminated by considering
the ratio of the number of strange quark pairs to the entropy $s/S$. 
In the  QGP, the dominant entropy production 
occurs during  the initial glue thermalization $\gamma_{\rm G}\to 1$, and the  thermal
strangeness production occurs in parallel and/or just a short time later.  
Moreover, both strangeness $s$ and entropy $S$ are nearly conserved in hadronization,
and thus, the final state  yield   value for the ratio $s/S$ is directly
related to the kinetic  processes  in the fireball at $\tau\simeq 1$--3 fm/c. 
A  thorough discussion of the observable $s/S$ is presented in~\cite{Kuznetsova:2006bh}, and 
detailed evaluation within a dynamical model of $s/S$ was  
obtained~\cite{Letessier:2006wn}. The following is a motivating introduction to these developments.

We first estimate the magnitude of $s/S$ in the QGP phase considering,  
in the hot  early stage of the reaction, an   equilibrated 
non-interacting QGP phase with perturbative properties:
\begin{eqnarray} 
{s \over S}\equiv\frac{\rho_{\rm s}}{\sigma}    
&=& 
\frac{\gamma_s^{\rm QGP} (3/\pi^2) T^3 (m_{  s}/T)^2K_2(m_{  s}/T)}
  {(32\pi^2/ 45)  T^3 
    +n_{\rm f}[(7\pi^2/ 15) T^3 + \mu_q^2T]},\nonumber \\
&=&  \label{sdivS}
{0.03\gamma_s^{\rm QGP}\over {1+ 0.054 (\ln \lambda_q)^2} }\,.
\end{eqnarray} 
Here, we used for the number of flavors $n_{\rm f}=2.5$ and $m_{  s}/T=1$. We 
see that the result is a slowly changing function  of $\lambda_q$,
for large $\lambda_q\simeq 4$ we find at lowest SPS energies, the 
value of $s/S$ is reduced by 10\%. Considering 
the slow dependence on $x=m_{  s}/T\simeq 1$ of $W(x)=x^2 K_2(x)$, there is 
further  dependence on the   temperature $T$. 

The rise with 
reaction energy toward the limiting value, $s/S=0.03$  for large 
$\sqrt{s_{\rm NN}}$,  is driven by the decrease in $\lambda_q\to 1$ and,
importantly, by an increase in chemical 
strangeness equilibration with the QGP occupancy $\gamma_s^{\rm QGP}\to 1$.
The dependence on the degree of chemical equilibration 
which dominates the functional behavior with $\sqrt{s_{\rm NN}}$ is:
\begin{equation}\label{sdivS2}
{s \over S}\!=\! { {0.03 \gamma_s^{\rm QGP}} \over 
  {0.38 \gamma_{\rm G}\!+ 
         0.12 \gamma_s^{\rm QGP}\!\!+
         0.5\gamma_q^{\rm QGP}\!\! + 
         0.054 \gamma_q^{\rm QGP} (\ln \lambda_q)^2}}.
\end{equation}

Eq.\,(\ref{sdivS2}) predicts  a smooth  increase in $s/S$ toward its
maximum value which by counting the degrees of freedom appears 
to be $s/S\to 0.03$, while the QGP source of 
particles approaches chemical equilibrium with increasing 
collision energy   and/or  increasing volume. 
It is important to keep in mind 
that the ratio $s/S$ is established early on in the 
reaction, and   the above relations and associated 
chemical conditions  we considered apply to
 the early hot  phase of the fireball. Yet, 
strangeness and entropy, once created, cannot disappear 
as the more complex low temperature domain is developing. 
Specifically near to   hadron freeze-out, the perturbative
QGP picture used above does not apply. Gluons are likely
to freeze faster than quarks and both are subject to much
more complex non-perturbative behavior. But the 
value $s/S$ is preserved across this non-perturbative domain.

\begin{table*}[!t]
\caption{
\label{AGSSPSse}
 AGS and CERN energy range 
(see top lines for projectile energy $E$ [GeV] and $\sqrt{s_{\rm NN}}$):
Strangeness yield $s$ ($=\bar s$), strangeness per entropy $s/S$, strangeness per baryon $s/b$,
the energy cost to make strangeness pair $\sqrt{s_{\rm NN}}/(2s/b)$,
thermal energy per baryon at hadronization $E_{\rm th}/b$, fraction of initial collision energy in  thermal
degrees of freedom, $(2E_{\rm th}/b)/\sqrt{s_{\rm NN}}$. 
}\vspace*{0.2cm}
\begin{center}
\begin{tabular}{|c| c | c c c c c |  }
\hline
E [$A$GeV]                   & 11.6        & 20      & 30         & 40         & 80            & 158  \\
$\sqrt{s_{\rm NN}}$  [GeV]  &4.84          &6.26     &7.61        &8.76        &12.32          &17.27  \\
\hline
\hline
$b $                        & 375.5        &347.9    &349.2       &349.9        &350.3        & 362.0      \\
$\bar s $                   & 35.5         &70.3     &100.8       &110          &161         & 218       \\
 \hline
100$\bar s/S$               &0.788         &1.26     &1.94        &1.90         &2.16          & 2.22       \\
$\bar s/b$                  &0.095         &0.202    &0.289       &0.314        &0.459         & 0.60       \\
 \hline
$\sqrt{s_{\rm NN}}/(2\bar s/b)$ [GeV] &25,5&15.5     &13.1        &13.9         &13.4         &14.4   \\
 \hline
$E_{\rm th}/b {\rm\ [GeV]}$  &1.82         &2.26     &2.33        &2.58         &3.30         & 4.08        \\
$(2E_{\rm th}/b)/\sqrt{s_{\rm NN}}$&0.752  & 0.722   &0.612       &0.589        &0.536        & 0.472     \\
$E_{\rm th}/\bar s {\rm\ [GeV]}$   &19.25  &10.9     &8.08        &8.21         &7.19        &6.80  \\
  \hline
\end{tabular}\vspace*{0.1cm}
\end{center}
 \end{table*}
 
\begin{table*}[!t]
\caption{
\label{Rhicse}
Top section: SHM yields of baryon $b$ and at central rapidity $db/dy$, 
and strangeness $s$ and $ds/dy$ at RHIC, left
for the total system, right for the central rapidity region. Next, we give 
strangeness per entropy $s/S$ (for central rapidity: $ds/dS$), strangeness per baryon $s/b$,
the energy cost to make strangeness pair $\sqrt{s_{\rm NN}}/(2ds/db)$,
thermal energy per baryon at hadronization $dE_{\rm th}/db$, 
fraction of initial collision energy in  thermal
degrees of freedom, $(2E_{\rm th}/b)/\sqrt{s_{\rm NN}}$. All  62.4 GeV results, and the 130 GeV $4\pi$ results,
 are,  as discussed in text, result of assumptions, and/or interpolation  of 
yields, and/or parameters, and hence are a  prediction.
}\vspace*{0.2cm}
\begin{center}
\begin{tabular}{|c| c  c c | c c c |  }
\hline
$\sqrt{s_{\rm NN}}$  [GeV] & 62.4 & 130  &200 & 62.4 & 130 & 200  \\
\hline
 &  \multicolumn{3}{c}{  $N_{4\pi}$ } & \multicolumn{3}{|c|}{ $dN/dy|_{y=0}$ }\\
 \hline
$b$, $db/dy$                           &350  &350     &350   &33.5   &18.5   &14.8        \\
$\bar s$, $d\bar s/dy$                 &560  &755     &726   &120.4  &136.7  &123         \\
\hline
100$\bar s/S,\ d\bar s/dS$             &3.17 &2.43    &2.66  &3.30   &3.39   & 3.13        \\
$\bar s/b,\ d\bar s/db$                &1.60 &2.16    &2.07  &3.35   & 6.87  & 8.29        \\
 \hline
$\sqrt{s_{\rm NN}}/(2d\bar s/db)$ [GeV]&19.5 &30.1    &48.3  &9.31   &9.46   &12.06         \\
 \hline
$dE_{\rm th}/db {\rm\ [GeV]}$          &7.18 &9.52    &11.24 &14.8   &30.4   & 38.2     \\
$(2dE_{\rm th}/db)/\sqrt{s_{\rm NN}}$  &0.230& 0.146  &0.112 &0.474  &0.467  & 0.382     \\
$dE_{\rm th}/d\bar s {\rm\ [GeV]}$     &4.49 &4.41    &5.41  &4.42   &4.42   &4.60    \\
 \hline
\end{tabular}\vspace*{0.1cm}
\end{center}
 \end{table*}

\begin{figure}[!tb]
\epsfig{width=9.5cm,figure=\pathnow  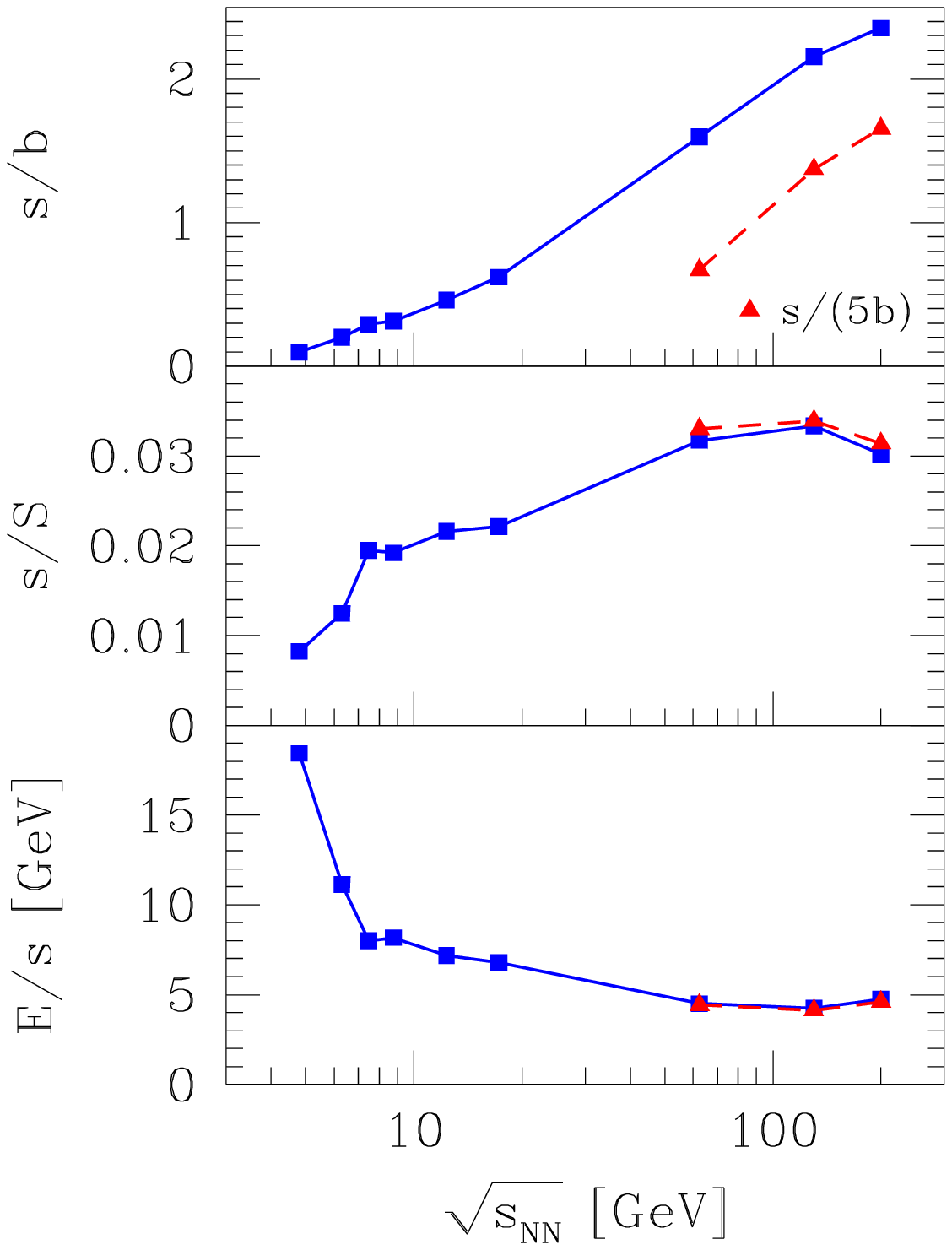}
\vskip -0.5cm
\caption{\label{sSb}
The specific strangeness yield as function of reaction energy 
$\sqrt{s_{\rm NN}}$. Top strangeness per baryon $\bar s/b$, middle 
 strangeness per entropy $\bar s/S$  and 
bottom $E_{\rm th}/\bar s$ thermal energy content per strangeness.
 Solid  squares correspond to $N_{4\pi}$
 the triangles on right are for the rapidity density yield $dN/dy$
at RHIC. The total yield results
are  connected by solid line to guide the eye, and the central rapidity
results (red) are connected by dashed line. 
}
\end{figure}

In  tables \ref{AGSSPSse} and \ref{Rhicse}, we present, in top portion, 
the   strangeness production as function of reaction
energy at AGS, SPS and RHIC, respectively.  We give 
the baryon content and  the total strangeness content of the fireball 
 derived  from the SHM fit to particle yield. 
Below, we see the above discussed  
strangeness per entropy $s/S$ ratio,  and strangeness per net 
baryon number  $s/b$ ratio.  
We present the increasing specific strangeness per baryon and per entropy
yields in figure \ref{sSb}, two top panels. The remarkable result we find is that the 
specific per entropy yield of strangeness converges for top RHIC 
energy and central rapidity toward the QGP result obtained counting 
the degrees of freedom, see Eq.\,(\ref{sdivS}). The somewhat smaller 
values for the $4\pi$ case are consistent with the average being made over 
the fragmentation region. This effect is greater in the ratio $s/b$ as we
have to count  all participant baryons.

The   middle section,  in  table \ref{AGSSPS}, 
 shows the center of momentum energy cost $\sqrt{s_{\rm NN}}/(2s/b) $ to make 
one   strangeness pair. The micro canonical input variables, $s/V$ and $b/V$, for this entry   
vary  significantly along with, and  as function of  $\sqrt{s_{\rm NN}}$. Yet, 
we see that the result obtained varies smoothly, at first it diminishes 
finding a minimum at around $E=40\,A$\,GeV and it rises slowly thereafter.
It is clearly more energy expansive to
make strangeness at AGS, nearly by factor 2. A minimum in energy cost 
to make strangeness is near to 30 $A$GeV beam energy, 
at the peak of ${\rm K}^+/\pi^+$ horn.

The increase in cost of making strangeness 
can be  attributed to the decreasing energy fraction 
stopped in  the reaction. 
 The energy stopping can be estimated by evaluating 
the per baryon thermal energy content $E_{\rm th}/b$ and obtaining from this 
the fraction of the initial energy converted into thermal energy
in the final state,  $(2E_{\rm th}/b)/$ $\sqrt{s_{\rm NN}}$,
 which fractions steadily drops from 75\%  at AGS to 48\% at 
top SPS energy. 

In terms of thermal energy, the cost of making strangeness
pair is given in the last line of  table \ref{AGSSPS}. After
an initial very rapid drop from AGS cost at $20$ GeV to 8 GeV near to
the top of the horn, there follows  a very slow and gradual decrease.
We show this   result graphically in the bottom panel in figure \ref{sSb}.
This behavior clearly shows a {\it rapid}  but smooth change-over  in the underlying 
mechanism of  strangeness production with  increasing 
 reaction energy,   between  11.6 and 30 $A$GeV. Once the new mechanism
is fully operational, we have essentially a flat, slowly decreasing 
 energy cost   per strangeness.
The drop we observe above  30 $A$GeV can be thought to originate in
transfer of thermal energy to the kinetic energy of collective expansion
which we do not record in our analysis, and thus, it is conceivable that
the cost in actual energy remains constant  above $\sqrt{s_{\rm NN}^{\rm cr}}$.  

As the bottom right of table \ref{Rhicse} indicates, the fraction
of energy stopped in the central rapidity  region at RHIC,
$(2dE_{\rm th}/db)/\sqrt{s_{\rm NN}}$ is 
rather large, it is estimated to be 58\% at $\sqrt{s_{\rm NN}}=62.4$ GeV
decreasing to 36\% at top RHIC energy. The energy cost
to make strangeness extrapolates well from  the SPS  level, connecting smoothly, 
see the bottom panel in figure \ref{sSb}, for both total yield and central rapidity yield. 
We note, in passing, that
only a small fraction, 10\%,  of the total energy is   thermalized at
the top RHIC energy considering the total fireball. 90\% is evidently
the energy of the   collective flow, predominantly in the longitudinal 
direction.

\begin{figure}[!tb]
\hspace*{-.40cm}\epsfig{width=10.2cm,figure=\pathnow    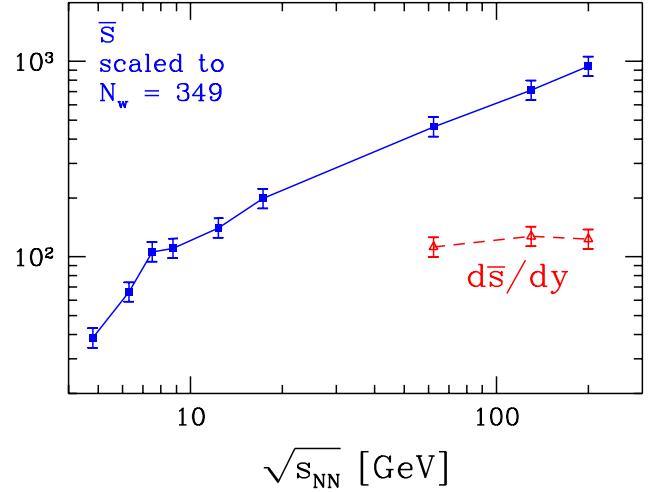}
\vskip -0.5cm
\caption{\label{syield}
The strangeness $\bar s$ (= $s$) content resulting from the SHM fit,
as function of reaction energy. The total yield results, solid squares (blue), are 
scaled with number of wounded nucleons to a fireball formed in 7\% central Pb--Pb
reactions $(N_{\rm w}=349)$. The triangles, on right, are 
for the rapidity density yield $d\bar s/dy$ at RHIC.
}
\end{figure}

The expectation of ever rising strangeness yield with 
$\sqrt{s_{\rm NN}}$ are not disappointed in Fig. \ref{syield}, 
but the  rapid smooth rise is surprising.   One finds such a result 
in a nearly model independent analysis adding up the 
$\bar s$ carrier particles, which are mostly directly
measured. A more 
precise study which adds up strangeness in the particles produced 
according to the SHM as seen in tables \ref{AGSSPSse} and \ref{Rhicse} 
is shown in  figure \ref{syield} --- 
there are non-negligible contributions of unobserved hidden strangeness,
in particular  in the $\eta$ hadron (40\% $s\bar s$ content). We have scaled the strangeness
yield to the 7\% centrality with $N_{\rm W}=349$ for the total yields.
For the central rapidity, we present results for the 5\% centrality. 

\section{Discussion and Interpretation} \label{results}
\subsection{The K$^+/\pi^+$ horn}\label{horn}
One can wonder how, in qualitative terms, 
can a parameter $\gamma_q$, which controls the light
quark yield, help explain   the horn structure seen in  figure  \ref{KPi}. 
We observe that this   horn structure  in the ${\rm K}^+/\pi^+$ ratio  
traces the final state  valance quark ratio $\bar s/\bar d$, and in language of
quark phase space occupancies $\gamma_i$ and fugacities $\lambda_i$, we have: 
\begin{equation}\label{Kpi}
{{\rm K}^+\over \pi^+}\to {\bar s\over \bar d} \propto F(T)
  \left({\lambda_s\over \lambda_d}\right)^{\!-1} {\gamma_s\over \gamma_d}
  \simeq 
   F(T)\left(\lambda_{I3}{\lambda_s\over \lambda_q}\right)^{\!-1} 
                                   {\gamma_s\over \gamma_q}.
\end{equation}

In chemical equilibrium models $\gamma_s/\gamma_q=1$, and the horn effect
must arise solely from the variation in the ratio $\lambda_s/ \lambda_q$ and the 
change in temperature $T$. The isospin factor $\lambda_{I3}$ is insignificant in
this consideration. For the interesting 
range of freeze-out temperature, $F(T)$ is a smooth  function of $T$. Normally,
one expects that $T$ increases with collision energy, hence we expect an
monotonic  increase  in the ${\rm K}^+/\pi^+$ ratio,  not considering the 
  quark chemistry. 

As collision energy is increased, increased hadron yield leads to 
  a decreasing $\lambda_q=e^{\mu_{\rm B}/3T}$.
We recall the smooth decrease of $\mu_{\rm B}$ with 
reaction energy   seen in bottom panel in figure \ref{gammu}.
The two chemical fugacities  $\lambda_s$ and $\lambda_q$ are
coupled by the condition that the strangeness is conserved. The chemical
potential effect is suggesting a smooth increase in the K$^+/\pi^+$ ratio. 
With considerable effort, one can arrange the chemical equilibrium fits to 
bend over to a flat behavior at  $\sqrt{s_{\rm NN}^{\rm cr}}$ as the dotted line
in figure  \ref{KPi}  shows.   It is quasi impossible to generate a horn 
with chemical equilibrium model.

Consideration of chemical non-equilibrium allows us 
to consider an energy dependent   ratio $\gamma_s/\gamma_q$,  which as
seen in Eq.\,(\ref{Kpi}) is a multiplicative factor in  the horn structure.
The fit produces    a horn like behavior of  $\gamma_s/\gamma_q$
at  $\sqrt{s_{\rm NN}^{\rm cr}}$, seen in
figure \ref{gammu}.   
As function of energy, many  other particle yields must
remain relatively smooth, with a few exceptions seen 
in figure \ref{LamAl}. We see that the 
description of the horn structure is possible, as there are 
effectively three function of  $\sqrt{s_{\rm NN}}$ 
which help to create it, $T$, $\lambda_q/\lambda_s$ and $\gamma_s/\gamma_q$,
but it is     in no way assured that the right horn arises, 
seen the behavior with energy  of the other particle yields. 

Indeed,
only the full chemical non-equilibrium model in which the 
two phase space occupancies,  $\gamma_s$ and $ \gamma_q$, vary independently,
does a good job as is seen comparing the solid with dashed  and dotted lines in 
 figure~\ref{KPi}.  Seen the horn-like  structure of all these lines obtained relaxing 
strangeness conservation constraint we realize that it is  not   the increased 
number of parameters, but the fact that particle production follows
the   SHM  with chemical non-equilibrium which allows the non-equilibrium model  to succeed.

\subsection{The K$^+/\pi^+$ horn as function of reaction volume}\label{Univ}

The rather sudden changes in 
freeze-out parameters $\gamma_q$  and $T$ appears to be
a  universal behavior. 
We established  it here as function of energy, and in earlier 
work as function    of  the   reaction volume ({i.e.}, participant number $A$),  
see figure 1 in~\cite{RafRHIC}. In both cases, 
  the chemical freeze-out temperature is higher  below a  threshold,
as expressed either by low 
energy or participant number. The most
drastic change is that   $\gamma_q$ jumps up from a value at, 
or below 0.5, to 1.6 as either 
the energy or  volume threshold is crossed. The   volume 
threshold is, however, not as sharp as the reaction energy 
threshold. The large system limit is achieved for $A> 25$, with a 
smooth transition beginning at $A> 6$, as can be seen in 
 figure 4 in \cite{RafRHIC}. 

Seeing this remark,
one immediately wonders if the ${\rm K}^+/\pi^+$ horn is present in the 
impact parameter kaon and pion data and the answer is no.  
Actually, this is not
surprising: since both $\pi^+$ and ${\rm K}^+$ originate,  in our study, 
at the level of  about 50\% in directly thermally produced  particles
the ratio ${\rm K}^+/\pi^+ $ is a measure of
the horn structure is  due to a rise in density of strangeness 
$\bar s$ at hadronization, outpaced   by the rise in the
$\bar d$ density  above  $\sqrt{s_{\rm NN}^{\rm cr}}$, whatever the mechanism in
terms of statistical parameters that implements this. However, when 
considering the impact parameter dependence at  $\sqrt{s_{\rm NN} }=200$ GeV,
the rise in strangeness has yet to occur, as in the small 
volume there has insufficient life span to produce strangeness. In this situation
we do not expect that the   horn is present as function of $A$.

\begin{figure}[!tb]
\hspace*{-.20cm}\epsfig{width=9cm,figure=\pathnow   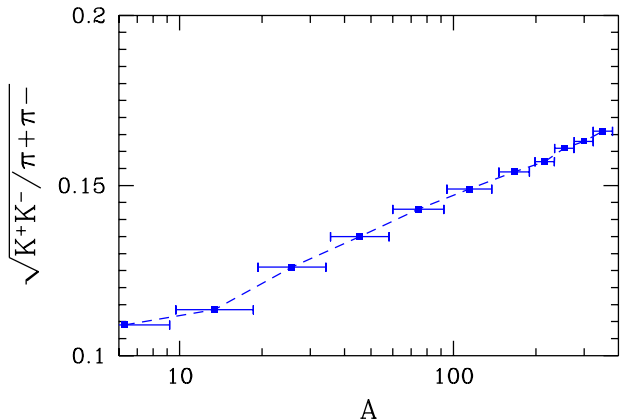}
\vskip -0.6cm
\caption{\label{KKPiPi}
  $ \sqrt{{\rm K}^+{\rm K}^-/\pi^+\pi^-}$ as function of 
participant number $A$ varying with   reaction centrality, 
PHENIX data~\cite{phenixyield}.
}
\end{figure}

One can  see the delayed production of strangeness as function of 
impact parameter directly in the PHENIX impact parameter
data~\cite{phenixyield}, without need for a detailed theoretical
 analysis. Consider  the ratio shown in figure \ref{KKPiPi}:
$K/\pi\equiv \sqrt{{\rm K}^+{\rm K}^-/\pi^+\pi^-}$.
This particular product-ratio of particles is nearly
 independent of chemical potentials
$\mu_{\rm B}, \mu_{\rm S}$ and the volume $V$ since
 it comprises ratio  of products of
particles and antiparticles. The rise seen in figure \ref{KKPiPi} is 
 evidence for an additional strangeness production mechanism turning 
on at about $A\simeq 20$. In figure \ref{KKPiPi}
we do not show a common systematic error, thus the normalization scale of the figure
could undergo a revision. This cannot change the insight that the additional 
strangeness above and beyond the first collision content is produced 
for $A>20$, enhancing the global yield
by 50\% or more. Moreover, we see that the rise 
is gradual as can be expected in 
kinetic theory models of strangeness 
production~\cite{Letessier:1996ad,Letessier:2006wn}, and there is at the maximum
centrality no evidence as yet of strangeness yield saturation.

The entropy content of the small system $A<20$ is such that 
strangeness per entropy  is at the level of $s/S\simeq 0.02$, and both
entropy and strangeness rise 
with centrality of the reaction at  $\sqrt{s_{\rm NN} }=200$ GeV.
However, unlike the energy dependence, the ratio $s/S$ rises   modestly, 
strangeness does not outpace entropy rise by more than 20\%. This is 
in agreement with expectation, since the threshold of strangeness mass is
not fully relevant at the top RHIC reaction energy, 
and thus we are seeing the properties of 
a deconfined initial state in which strange quark is effectively massless. 
Instead, it is the lifespan of the system that matters, as noted above.
 
There is very little observed   dependence of ratios 
of hadron resonances with the
ground state yields, such as $K^*/K$. This implies and agrees in 
quantitative way with the tacit assumption inherent in the above discussion,
and the result of a more detailed  analysis~\cite{RafRHIC},  
that there is no $T$ dependence of the
freeze-out conditions for  $A > 20$. For this reason for  $A>20$
ratios of all hadrons
which do not involve a difference in strangeness content,
do not vary with centrality.

We further note that there is little change in chemical potentials with 
centrality for $A>20$, indicating that the stopping of baryons is not a
result of multiple scattering, but is due to phase conditions of matter.
Comparing  other properties of matter,  we see very much the same behavior 
as function of  impact parameter and reaction energy: in particular, 
we note the step up in pressure, in 
energy density, and in entropy density 
at the impact parameter threshold~\cite{RafRHIC}. 
 
\subsection{Chemical equilibrium or non-equilibrium?}\label{equi}
 
An important questions discussed in the
 study of hadron yields interpretations 
is if chemical equilibrium or non-equilibrium 
prevails in the hadronization 
process. There are workers who strongly defend the
chemical equilibrium SHM~\cite{Braun-Munzinger:2003zd}. 
Let us look again at the survey of fit quality 
results seen in figure \ref{ChiP}. We
  note that for  $\gamma_q=1$ (but $\gamma_s\ne 1$)
at each energy   there seems to exist a 
reasonable fit with   $0.5<\chi^2/{\rm d.o.f.}<1.5$ for the data sample
considered,
which suggests that  at each reaction  energy with  $\gamma_q=1$
a reasonable and widely accepted physical description of the experimental data
  emerges. This result  is therefore claimed  in studies that  focus on the hadron yields 
at each energy separately. What works poorly in  SHM used with $\gamma_q=1$
and even worse with $\gamma_q=1,\ \gamma_s=1$
is the   energy dependence of  particle ratios, with the most
prominent present day  example being the   horn structure
 in  the  ${\rm K}^+/\pi^+$ yield ratio. Seen from this perspective,
it is the  energy dependent particle
 yield  that requires the inclusion  in the necessary set
of parameters  a  varying value $\gamma_q\ne 1$. 

Another important question directly related to the issue
of chemical equilibrium and non-equilibrium   is how the 
fitted results for $T(\mu_{\rm B})$, the `hadronization 
curve' relate to the phase boundary between deconfined 
primary phase and the hadron phase. Clearly, the result of
the fit are greatly dependent on the assumption about chemical
condition with the equilibrium fit claiming a hadronization 
at RHIC at $T=175$ MeV. 

The rapidly decreasing freeze-out 
temperature $T$ as $\sqrt{s_{\rm NN}}$ decreases, and which 
is certainly inconsistent with the rather flat phase transition
boundary at moderate chemical potentials is explained by suggesting 
that  the hadronization
may be related to a  particular values of energy per particle content,
of the magnitude 1 GeV ~\cite{Braun-Munzinger:2003zd}. 
However, this condition, though not  
 rooted in any known basic physical principle, is also
 obtained in some dynamical studies,
 see, {\it e.g.,} Ref. \cite{Bleicher:2002dm,Bratkovskaya:2004kv}. 
We note that the chemical 
equilibrium hypothesis fails to explain the hadronization conditions 
expected as function of $T$ and $\mu_{\rm B}$, or equivalently, as
function of $\sqrt{s_{\rm NN}}$.

In summary, the interpretation of hadron production in terms
of chemical equilibrium SHM disagrees, in quantitative manner, 
 both with  the reaction  energy dependent particle yields 
(such as the K$^+/\pi^+$  horn) and the reaction energy 
dependent shape of hadronization boundary.

\subsection{Hadronization boundary in heavy ion collisions}\label{bound}
We believe that the hadronization boundary,
in the $T$--$\mu_{\rm B}$ plane,  is the result
of a complex interplay between the dynamics of 
heavy ion reaction and the properties of both phases of
matter, the inside of the fireball,  and the hadron phase we
observe. Even disregarding complications related to  the 
rapid expansion of the dense matter fireball, the
presence of chemical non-equilibrium particle 
distributions introduces significant freedom into 
the shape and location of the  $T(\mu_{\rm B})$ transition region.

Recall, first, that  available  lattice results  apply to 
a system in the thermodynamic limit with $\gamma_q=\gamma_s$~=~1,
for both quark and confined hadron phases.
The typical boundary between the QGP and hadron phases is discussed 
in  Ref.\,\cite{Fodor:2004nz}, and is dependent on chemical properties of 
QGP. Typically, one considers the dependence on chemical potentials,
and in particular on $\mu_{\rm B}$, however, a  significant 
change  in the phase boundary location is 
to be expected when  $\gamma_q$ and $\gamma_s\ne$~1.
To understand this important remark, consider
the two other  known cases  
$\gamma_q=1,\ \gamma_s=0$ corresponding to 2 flavors, 
and $\gamma_q=\gamma_s=0$ corresponding to pure gauge. 
There is a significant change in $T(\mu_{\rm B}=0)$, which increases
with decreasing $\gamma_i$.

Moreover, not only the location
but also the {\it nature} of the phase boundary can be  modified by 
variation  of $\gamma_i$. We recall that for the 2+1 flavor case,
there is possibly a critical point at finite baryochemical potential
with $\mu_{\rm B}\simeq 350$ MeV~\cite{Fodor:2004nz,Allton:2005gk}. 
However, for the case of 3 massless flavors 
there can be  a 1st order transition 
at all $\mu_{\rm B}$ \cite{Peikert:1998jz,Bernard:2004je}. 
Considering a classical particle system, one easily sees 
that  an  over-saturated
 phase space, e.g., with  $\gamma_q=1.6,\ \gamma_s\ge \gamma_q $  
for the purpose of the study of the phase transition acts   
 as being equivalent to a system with 
3.2 light quarks and 1.6 massive (strange)
quarks present in the confined hadron phase. 

Even though one should be keenly aware that 
over-saturation of the phase space is not the same 
as additional degeneracy due to true degrees of
freedom, the similarity of resulting effect must be considerable. 
We know that with increasing $\mu_{\rm B}$, the increased
quark density creates the environment in which the 
phase cross-over becomes a phase transition. 
The influence of $\gamma^{\rm QGP}_{q,s}$ cannot be 
different. Considering that  $\gamma^{\rm QGP}_{q,s}$ 
enhances both quark and  antiquark number, it should be 
more effective compared  to $\mu_{\rm B}$ in its 
facilitation of a  phase transition, and reduction of 
the temperature of the phase boundary for $\gamma^{\rm QGP}_{q,s}>1$.

We therefore can expect that,  for a chemically over-saturated system, 
there is also an  effective  increase in the
  number of degrees of freedom. 
Looking at the structure of the quark-hadron transformation this increase 
in the number of available effective
degrees of freedom occurs in a physical system which is 
almost, but not quite,   able 
to undergo a 1st order  phase transition.  Considering here also 
the sudden nature of the fireball breakup 
seen in several observables~\cite{RBRC}, 
we conjecture that the hadronizing fireball leading to
 $\gamma_s>\gamma_q=1.6$  passes
 a true  phase boundary   corresponding to a 1st 
order phase transition  condition  at small 
$\mu_{\rm B}$. Because of the changed count in the
 degrees of freedom, we expect that the phase transition temperature 
is at the same time decreased to below  the cross-over  value
for chemical equilibrium case of 2+1 flavors  near $T=162\pm3$ MeV.  
 
It seems to us that it would be very interesting to determine, in as
more  rigorous  way  for   the case of  
 the 2+1 flavor lattice QCD at $\mu_{\rm B}=0$ 
for which  values, if any, of $\gamma_i$
the   system undergoes a phase transition of 1st order.  
Lattice QCD methods employed to obtain 
results at finite $\mu_{\rm B}$, {\it e.g.}, the power 
expansion~\cite{Allton:2005gk,Allton:2003vx}, 
should also allow to study the case of 
   $\mu_\gamma\equiv T\ln \gamma_i > 0$, and
near to $\mu_\gamma=$, {\it i.e.} $\gamma_i=1$.
We see the actual difficulty in the need to simulate
different values of  $\mu_\gamma$ in the 
two phases, such that the quark pair
content is preserved across the phase boundary.

 The   dynamical,  and theoretically less spectacular, effect  
capable to shift the location in temperature of the 
expected phase boundary, is due to 
the expansion dynamics of the fireball. The analysis of the 
RHIC results suggests that    the collective 
flow occurs at parton level~\cite{Huang:2005nd}.
Collective flow of color (partons) is  like a wind capable 
to  push  out the color non transparent `true' vacuum~\cite{Csorgo:2002kt},
adding to thermal pressure the  dynamical component, for 
a finite expanding system this would lead to supercooling~\cite{suddenPRL}. This
dynamical effect will  push the hadronization condition to
lower   local freeze-out $T$ at high   $\sqrt{s_{\rm NN}}$, thus
flattening the boundary between the phases as function of $\mu_{\rm B}$. 
 In the context of results 
we have obtained, it is the smoothness of the ratio $P/\epsilon$ obtained
at hadronization   which   supports the possible relevance of dynamic
 phase boundary displacement. 
This behavior suggests a smoothly changing  dynamical break up 
 condition, potentially related to (hydrodynamic) flow.

\subsection{Our hadronization boundary and its interpretation}\label{ourbound}
The above two effects, the change in the location of the 
static phase boundary in presence  of chemical {\it non}-equili\-brium 
and the dynamics of collective matter flow toward breakup condition,
  are both non-negligible but hard to evaluate
quantitatively. We believe that they can  
 explain why the chemical freeze-out conditions
$T$ and $\mu_{\rm B}$ are as presented in  figure \ref{gammu}. Of particular
relevance is the low value of $T$ at high reaction energy, and relatively 
high value of $T$ at low reaction energy, just opposite what one finds 
 to be result of SHM analysis when chemical equilibrium is assumed.  

\begin{figure}[!tb]
\hspace*{-.60cm}\epsfig{width=9.7cm,figure=\pathnow     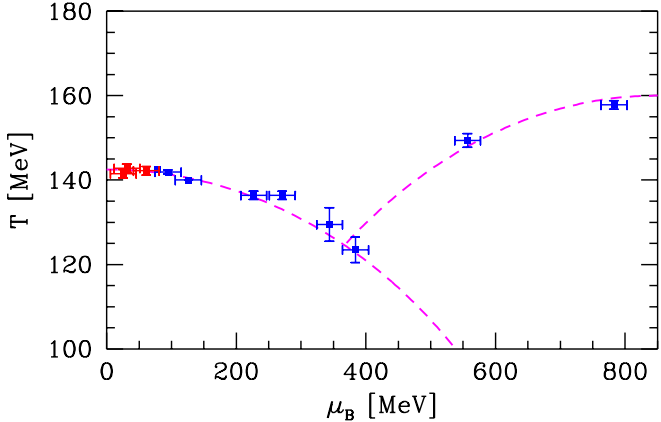}
\caption{\label{Tmu}
$T$--$\mu_{\rm B}$ plane with points obtained in the SHM fit. 
See text for discussion.   
}
\end{figure}

Our   chemical freeze-out  conditions are better  shown   in the 
$T$--$\mu_{\rm B}$ plane, see     figure \ref{Tmu}. Considering results
 shown in  figure  \ref{gammu}, we are able to assign to 
each point in the $T$--$\mu_{\rm B}$ plane the associate value of 
 $\sqrt{s_{\rm NN}}$. The RHIC $dN/dy$ results are to outer left.
They are followed by RHIC and SPS $N_{4\pi}$ results. The dip
corresponds to the 30 and 40 $A$GeV SPS results. The top right is 
the lowest  20 $A$GeV SPS and top 11.6 $A$GeV AGS energy range.  We see that the  
chemical freeze-out temperature 
$T$ rises for the two lowest reaction energies  11.6 and 20 $A$ GeV
 to near the Hagedorn temperature,  $T=160$ MeV. Such phase structure
is discussed e.g. in the context of chiral quark pairing~\cite{Kitazawa:2001ft}.

 The size of error bars, in figure \ref{Tmu}, is output of  the 
fit process, and when it is rather large,
 it implies that the resulting $\chi^2$ profile was relatively flat,
 or/and  that there were
two neighboring god fit  minima.  To guide the eye, we have
added two lines connecting the fit results. 
As seen in figure \ref{Tmu}, at $\mu_{\rm B}=0$, we find that 
hadronization occurs at   $T=140$,  decreasing to  $T= 120$ MeV at 
$\mu_{\rm B}=400$ MeV. Along this line $\gamma_q>1.6$. As argued
in the previous subsection \ref{bound}, this line could be
 a  true 1st order phase boundary
 between quark  matter and an  over-saturated hadron phase. 

Two different interpretations  come to mind when we  attempt to
understand   the other branch in  figure \ref{Tmu}, the 
rise   from $T\simeq 120$ to 160 for  $\mu_{\rm B}>400$, which is
accompanied by a rather low hadron side phase space occupancy.
  Most `natural' is to presume
that the dissolution of color bonds did not occur 
in heavy ion collisions below 30 $A$ GeV, we are dealing with 
`conventional' hadron matter. 
The under-saturation occurs   since there was no time to make hadrons, 
i.e., chemical equilibration was not achieved in the
colliding hadron system before it breaks apart.

The fireball  break-up   at a  higher temperature
is a consistent freeze-out scenario for under-saturated 
hadron phase space 
considering the  kinetic scattering 
freeze-out condition. Given the greatly reduced particle density 
($\propto \gamma_q^n, n=3,2,1$)  a  high $T$  freeze-out for $\gamma_i<0.5$
is consistent. The nucleon density 
scales with $\gamma_q^3$ and pion density 
with $\gamma_q^2$. Therefore the meson--baryon and meson--meson
scattering length scales as $L\propto 1/\gamma^5$, and $L\propto 1/\gamma^4$,
respectively. This  implies that, as the system expands, it is able to 
freeze  out early at a higher $T$. 

On the other hand,  the volume size we   found, see  
  table \ref{AGSSPS},  is significantly larger at low reaction energies.
This implies that a scenario with pure hadron matter present is subject 
to a  quite different expansion history. This signals that 
a  standard picture of a conventional  hadron matter formation
at reaction energies below the transition point 
at $6.26\,{\rm GeV}<\sqrt{s_{\rm NN}^{\rm cr}}<7.61\,{\rm GeV}$
may not be the  valid  explanation of results of our analysis. 
Namely, if the reaction history since 
first contact is different for the two reaction energy ranges, one would expect that 
 the systematics of the final state entropy production, strangeness production,  
and  strange antibaryon production has a visible break at the critical point. 
What we have found is, instead, that
 these quantities  show a rather  smooth uninterrupted rise with reaction energy. 

This  means that the the initial conditions reached 
in the reaction  where, e.g.,  entropy and strangeness
are produced, is not undergoing a sudden change.  The change occurs at the end
near to the hadronization of matter. For this reason,  we see a change in 
particle yields (the horn),   statistical parameters jump, and the physical 
conditions at hadronization jump even more. 
The yields of   quantities which are driven 
by physics of the initial dense matter formation,
e.g., entropy content, strangeness content,
change smoothly with heavy ion reaction energy in the domain we explored. 
We are furthermore swayed away from the picture of the 
hadronic gas  being the  form of matter at breakup 
below $\sqrt{s_{\rm NN}^{\rm cr}}$ by the strange antibaryon
production systematics we discussed in figure~\ref{Figantibaryon}.

We are searching thus for an explanation in terms of a new phase of  matter
being involved in the hadronization process, but
clearly this cannot be the semi-perturbative quark--gluon plasma state. 
The  conceivable explanation of the fit result below  30 $A$ GeV 
is  presence, at the high baryon density 
arising at large $\mu_{\rm B}$,  of a constituent quark
 plasma~\cite{Roizen:2004cy}. Even if the perturbative QCD
quark phase is reached at high temperature, in expansion--cooling
the system encounters the valon (word derived from `valance' quark)
phase in which the   color quark
bonds are broken, but chiral symmetry restoration is not completed, 
with quarks of mass $m_{u,d}\simeq 340$ MeV and $m_s\simeq 500$ MeV
being the only active   degrees of freedom. This scenario is not inconsistent
with the finding on the lattice, that, for  $\mu_{\rm B}\to 0$ and in chemical 
equilibrium,  the chiral symmetry restoration coincides with 
deconfinement transition.

In a valon matter, even assuming  chemical equilibrium, the number of 
 quark pairs at temperature near to $T=160$ MeV would be rather small,  
given the high constituent quark  mass.
In breakup of this system, a relatively small $\gamma_i^{\rm HG}$ is 
achieved. Furthermore, since
the mass of these constituent quarks  is greater than that of the pion,   
the phase transformation between hadron and valon matter  
occurs at  relatively large  $T$. To see this, recall that   the 
pion  with its  low mass produces  greater pressure than valons and 
thus is   pushing  
the transition boundary to higher $T$. Strangeness, and importantly
the entropy content in
this phase arise  due to prior initial state  perturbative QGP phase and hence such 
a valon system must be larger in volume at the point of hadronization. 

It is also
conceivable that a hadron fireball evolving from the beginning and 
  fully in the  valon phase would
maintain much of the continuity we saw in hadronic observables. 
For example, $u$ and $d$-valon-quark scattering can produce strange
valon-quark  pairs, and these give rise in hadr\-on\-ization to the abundances
of strange antibaryons, as expected in the deconfined phase. What speaks for
this option is the rather sudden change in the   thermal  energy content
per strange quark pair produced,  which 
is seen in bottom of figure \ref{sSb},
 indicating appearance of a new energy  efficient mechanism 
of strangeness production above $\sqrt{s_{\rm NN}^{\rm cr}}$.

One may wonder how our findings compare to earlier studies of the 
phase boundary, both in statistical models~\cite{Becattini:2003wp},
 and microscopic models (see Ref.~\cite{Bratkovskaya:2004kv}  
and references therein).  In the microscopic models 
one  accomplishes  a better understanding of the approach
to thermal  and chemical equilibrium of the degrees of freedom  
employed. A continuous phase boundary is, here, a direct outcome of 
the assumption made about the degrees of freedom present. 
Our analysis, which does not rely on such assumptions, is thus  
less model dependent and allows for presence of  degrees of freedom
with  unexpected  properties. On the other hand, we also firmly believe, 
as is shown in  figure  \ref{Tmu}  
that there is a smooth phase boundary, with $T$ 
dropping with increasing $\mu_{\rm B}$. What our study has uncovered
is the possible presence of another    phase boundary 
for $\mu_{\rm B}>350 $ MeV  at higher $T$. It is important for the
reader to keep in mind that this   finding  is not in conflict with 
theoretical chemical equilibrium  results which focus on the 
other, conventional, phase branch and address physics of phase 
transformation  occurring in  the  early Universe.

Moreover, a recent study of the low energy
AGS pion production data~\cite{Klay:2003zf} found that the 
thermal freeze-out temperature at reaction energy of 8 GeV 
is at $T_{\rm th}=140$ MeV~\cite{Pin-zhen:2005ca}. 
Thus  there is also consistency of our present analysis
with the shape of pion transverse mass spectra and the
high chemical freeze-out temperature we find at AGS. Another
interesting finding was the medium mass modification
which allowed to describe pion decay spectra.

\subsection{Final remarks}\label{final}

In summary, we have performed a complete analysis of the energy dependence
of hadron production in heavy ion collisions, spanning the range beginning
at the top AGS energy, to the top RHIC energy. We have made
extensive predictions about particle production in the entire
energy range. These results are useful in several respects. 
For example,  we have shown that the 
best energy to search for the elusive pentaquarks would be at SPS
at 30--40 $A$ GeV, where we find that the total yield of  $\Theta^+(1540)$   is already fully
developed. Thus there is a maximum in the ratio   $\Theta^+/K_{\rm S}\simeq 0.2$ 
at 30 $A$ GeV. 
Of course, this finding presupposes the existence of the exotic state.

We have furthermore presented hadron yields important in the understanding 
of dilepton spectra, such as $\rho, \eta$ and $ \omega$. The relative meson resonance 
yields we find do not follow the $pp$ systematics and vary as function of energy. 
One thus can test the hadronization picture here presented in study of resonance production.
This observation was recently exploited in a systematic fashion~\cite{Torrieri:2006yb}.

We have shown that the threshold in energy which generates a 
horn in the ${\rm K}^+/\pi^+$ yield ratio can be associated
with the chemical freeze-out shifting rather rapidly 
toward condition of greatly increased hadronization densities. This  transition  
 separates the high
entropy density phase at high heavy ion reaction energy from a low entropy density
 phase. This behavior parallels the  findings 
for impact parameter dependence of RHIC results, where the 
low entropy density phase is seen for small reaction volumes present 
at large impact parameters~\cite{RafRHIC}.

Several observables, including 
  strangeness    production, show   continuity across the energy 
threshold at $6.26\,{\rm GeV}<\sqrt{s_{\rm NN}^{\rm cr}}<7.61\,{\rm GeV}$,
thus, it seems that the critical conditions expresses 
a change in the nature of the fireball breakup, and to a lesser degree
  a reaction energy  dependent  change in the nature of initial conditions 
reached in the reaction. 

We have discussed, in depth, our  findings about the hadronization condition
$T(\mu_{\rm B})$ and have argued that   at high reaction energies 
  a 1st-order phase transition is arising in the chemically 
non-equilibrated hot hadronic matter system.
Detailed discussion was presented about possible changes in
phases of hadronic matter as function of reaction energy and reaction volume.  

\begin{acknowledgement}
This manuscript nucl-th/0504028, on which this work is based,
has been first web-published in April, 2005. We undertook 
the current revision to correct an entropy yield error which the early SHARE release contained. 
This changes entropy $S$ related table entries and figures, the data fits are unaffected. When 
redoing the fits to obtain entropy, we incorporated 
the latest strange hadron yields of NA49 as available.
We thank NA49 and PHENIX collaborations for  valuable comments regarding the
acceptances of weak decays. We thank M. Ga\'zdzicki and G. Torrieri for 
 valuable comments.
Work supported in part by a grant from: the U.S. Department of
Energy  DE-FG02-04ER4131.
LPTHE, Univ.\,Paris 6 et 7 is: Unit\'e mixte de Recherche du CNRS, UMR7589.
\end{acknowledgement}

\vskip -0.3cm


\begin{thebibliography}{19}
\bibitem{Glendenning:1984ta}
N.~K.~Glendenning and J.~Rafelski,
Phys.\ Rev.\ C {\bf 31}, 823 (1985);
note that   in Fig. 2   the pion yield from 
resonance decays is not included, this 
 dilutes the expected yields by about
factor two, flattening the curve.
\end{thebibliography}
\end{document}